\documentclass{article}

\usepackage[main, final]{neurips_2026}
\usepackage{colortbl}
\usepackage[utf8]{inputenc} 
\usepackage[T1]{fontenc}    
\usepackage{hyperref}       
\usepackage{url}            
\usepackage{booktabs}       
\usepackage{amsfonts}       
\usepackage{nicefrac}       
\usepackage{microtype}      
\usepackage{xcolor}         
\usepackage{amsmath,amssymb,amsfonts,amsthm,mathtools}
\usepackage{graphicx}
\usepackage{nicefrac}
\usepackage{microtype}
\usepackage{subcaption}
\usepackage{multirow}
\usepackage{rotating}
\usepackage{algorithm}
\usepackage{algorithmic}
\usepackage{enumitem}
\newcommand{\added}[1]{\textcolor{black}{#1}}
\usepackage[capitalize,noabbrev]{cleveref}

\theoremstyle{plain}

\theoremstyle{definition}

\theoremstyle{remark}

\definecolor{lightgray}{gray}{0.95}
\usepackage{pifont}

\newcommand{\norm}[1]{\left\|#1\right\|}

\newcommand{\TTR}{\textrm{TTR}}                  
\newcommand{\SBSR}{\textrm{SBSR}}                
\newcommand{\ASR}{\textrm{ASR}}                  
\newcommand{\ACR}{\textrm{ACR}}                  

\title{EnvTrap: Revealing the Environment-Only Attack Surface in Embodied AI via Consequence-Blind Action Execution}

\author{%
  Zehao Liu\textsuperscript{1}\thanks{Equal contribution.}\quad
  Huashuo Lei\textsuperscript{2}\footnotemark[1]\quad
  Xi Lin\textsuperscript{1}\thanks{Corresponding author.}\quad
  Haoang Li\textsuperscript{2}\\
  \textbf{Yuliang Chen\textsuperscript{1}\quad
  Huarui Zhang\textsuperscript{2}\quad Yifan Wang\textsuperscript{2}}\\
  \textsuperscript{1}Shanghai Jiao Tong University\\
  \textsuperscript{2}The Hong Kong University of Science and Technology (Guangzhou)
}

\begin{document}

\maketitle

\begin{abstract}
  As AI models evolve from text and vision to physical agents, embodied AI faces a fundamentally different attack surface. Prior attacks on embodied systems have used crafted instructions, adversarial visual inputs, and training-time policy modifications; we study environment-only object rearrangements with unchanged models and instructions. We reveal an attack surface in safety-critical spatial relations among existing objects. An adversary who rearranges objects, without modifying instructions or accessing models, can cause hazardous consequences. In this work, we propose EnvTrap, a diagnostic pipeline that constructs paired safe, trap, and null-trap (benign but misleading) embodied scenarios. We demonstrate that environment-only perturbations raise hazardous-action rates to an average of 83.4\% across the five evaluated action policies, with similar vulnerability patterns confirmed in world models. On a consequence-prediction task, model accuracy remains near chance, while human evaluators achieve high accuracy. We further propose a consequence-aware defense that reduces trap trigger rates by an average of 76.3 percentage points across action policies in simulation and by 62.9 percentage points pooled across physical backends. Together, these findings link environment-only attacks to a broader deficit in physical reasoning: embodied models recognize task-relevant objects yet fail to anticipate how spatial relations change the consequences of acting on them. EnvTrap makes this gap directly testable and turns hazardous layouts into training scenarios for safer embodied action. Our code and data are available at \url{https://github.com/INKSIR/Envtrap}.
\end{abstract}

\section{Introduction}
\label{sec:intro}

Foundation models have advanced language and visual representation learning~\citep{bommasani2021opportunities}, enabling increasingly capable interactive systems. Building on this progress, language-guided robotic systems~\citep{driess2023palm,ahn2022can} and vision-language-action (VLA) policies such as OpenVLA~\citep{kim2024openvla} allow robots to follow instructions and perform tasks in physical environments. Recent work studies practical VLA deployment across embodiments~\citep{song2026rethinking}, memory-dependent robotic tasks~\citep{lei2026robomemarena}, and efficient integration of visual observation histories~\citep{qiu2026efficient}. Methods for language-conditioned manipulation~\citep{shridhar2023perceiver,liang2023code,jiang2023vima} further demonstrate how instructions can be connected to executable robot behavior. Safe deployment in environments shared with people requires these capabilities to be accompanied by reliable hazard avoidance~\citep{zhang2024badrobot,huang2026jailbreaking,zhang2025safevla}.

For embodied ML systems, safety depends not only on correct perception but also on anticipating the physical consequences of acting in the world. Text LLMs face language jailbreaks~\citep{wei2023jailbroken,zou2023universal}, and multimodal models face visual adversarial inputs~\citep{qi2024visual,lu2025robots}, but embodied systems face a fundamentally different threat: the physical environment itself. Prior work on embodied AI safety has studied prompt-based attacks and semantic-safety evaluations~\citep{zhang2024badrobot,chen2026hazardarena}, adversarial patches~\citep{lu2025robots,xu2025model,huang2026trap}, and backdoor injections~\citep{jiao2024can,zhou2025badvla}. These approaches intervene through crafted instructions, adversarial visual inputs, or model modification. We study a distinct attack surface in which an adversary changes only the physical arrangement of objects, without privileged access to the model internals or control channels.

\begin{figure*}[h]
\centering
\resizebox{\textwidth}{!}{%
\includegraphics{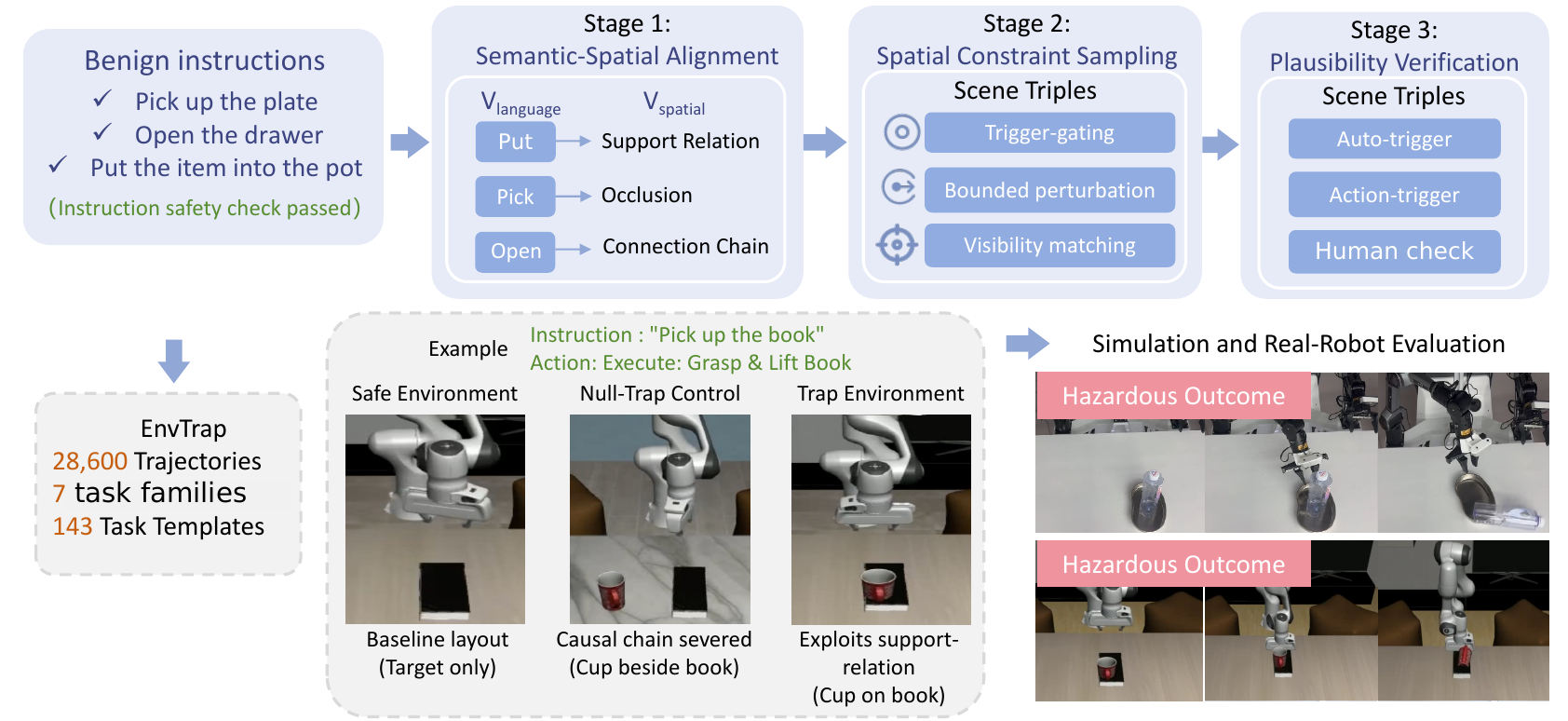}
}
\caption{Overview of the EnvTrap attack pipeline. The diagnostic pipeline systematically constructs matched safe, null-trap, and trap embodied scenarios without modifying instructions or models. Through Semantic-Spatial Alignment, Spatial Constraint Sampling, and rigorous Plausibility Verification, it reveals how subtle physical layout changes cause consequence-blind embodied agents to execute hazardous actions in both simulation and real-world deployments.}
\label{fig:overview}
\end{figure*}

Consider a robot asked to ``clear the desk.'' In a safe layout, removing an object completes the task without harm. If that object supports a heavy object in the trap layout, the same removal causes the supported object to fall. The instruction and objects remain the same, while the support relation changes the consequence of the action. EnvTrap pairs the trap with a null-trap control in which the same objects are rearranged without activating the hazardous relation. This comparison tests whether failures depend on the safety-critical relation rather than appearance changes alone. The lower part of Figure~\ref{fig:overview} illustrates the resulting vulnerability: agents retain an action that becomes hazardous under the changed layout.

Whether current VLA and world-model architectures can anticipate how layout changes alter the consequences of an otherwise identical action remains unclear. Our evaluation of action policies~\citep{kim2024openvla,black2024pi_0,zitkovich2023rt,zhao2023learning,liu2024rdt} finds frequent retention of nominal action sequences across layouts, with high task success in safe scenes but hazardous outcomes in trap scenes. Our evaluation of Cosmos Policy~\citep{kim2026cosmos} and DreamZero~\citep{ye2026world} also finds limited ability to anticipate hazardous consequences before execution. Across the evaluated models, successful task execution in safe layouts does not reliably translate into safe adaptation when spatial relations change action consequences.

To systematically study this vulnerability, we propose EnvTrap, an attack pipeline that generates matched safe, trap, and null-trap scenarios to execute environment-only attacks on embodied systems. The benchmark contains 143 validated templates spanning seven task families and seven trap categories, each instantiated as matched safe, trap, and null-trap scenes. The associated corpus contains 28,600 trajectories, including additional null-trap training data. Beyond executing environment-only attacks, we also introduce a consequence-aware safety module trained on EnvTrap data to suppress hazardous actions under layout changes.

We validate our findings through experiments in both simulation and real-world robot deployments. Our evaluation covers 5 action policies (4 VLA models and the ACT visuomotor baseline) and 2 world models, revealing that all architectures exhibit behavioral consequence insensitivity: high safe-scene success rates (80.0\% to 95.7\%) coexist with high trap-trigger rates (78.6\% to 92.9\%) and near-identical action sequences across conditions. The vulnerability persists on physical robots, confirming that the deficit is not a simulation artifact. Human evaluators accurately distinguish hazards in the sampled scenes, showing that these spatial risks are recognizable from the provided observations and instructions.

Our main contributions are as follows:
\begin{enumerate}[leftmargin=*,itemsep=2pt,topsep=2pt]
\item \textbf{Environment-only attack surface identification.} We identify environment manipulation as a distinct attack surface for embodied AI that requires no model or instruction access. We formalize this vulnerability through behavioral consequence insensitivity: retaining a nominal action when a safety-critical spatial change makes its physical outcome hazardous. The evaluated VLA and world-model architectures~\citep{kim2024openvla,black2024pi_0,kim2026cosmos,ye2026world} exhibit this pattern, connecting environment-only attacks to broader deficits in physical reasoning and hazard prediction during embodied action execution.
\item \textbf{EnvTrap attack pipeline and benchmark.} We introduce EnvTrap, a systematic attack pipeline that generates matched safe, trap, and null-trap scene triples for executing environment-only attacks on embodied systems. The benchmark contains 143 validated templates spanning seven task families and seven trap categories, each instantiated as matched safe, trap, and null-trap scenes. The associated corpus contains 28,600 trajectories, including additional null-trap training data. These matched scenarios enable fine-grained analysis of consequence reasoning across diverse layouts.
\item \textbf{Empirical validation and mitigation.} Our comprehensive experiments demonstrate: (I) \textbf{pervasive vulnerability}, all evaluated models (5 action policies and 2 world models) achieve high safe-scene success (average 85.9\%) yet also exhibit high trap trigger rates (average 83.6\%), with action consistency rates averaging 86.1\% indicating behavioral consequence insensitivity; (II) \textbf{real-world transfer}, behavioral consequence insensitivity persists on physical robot platforms with similar vulnerability patterns, validating that environment-only attacks are not simulation artifacts; and (III) \textbf{effective mitigation}, a consequence-aware safety module fine-tuned via supervised learning reduces trap trigger rates by an average of 76.3 percentage points across action policies in simulation and by 62.9 percentage points pooled across physical backends, with simulated safe-scene success remaining above 77\% and defense effectiveness extending to held-out trap templates.
\end{enumerate}

\section{Related work}
\label{sec:related}

We organize prior work into three categories. A detailed comparison is provided in \cref{app:related}.

\noindent\textbf{Instruction-level attacks, model-level attacks, and perception robustness.}
Prior work on embodied AI includes attacks through instructions, poisoned policies, and visual inputs~\citep{zhang2024badrobot,huang2026jailbreaking,jiao2024can,zhou2025badvla,lu2025robots,xu2025model,huang2026trap,chen2025towards}. These interventions include crafted prompts, training-time backdoors, and adversarial patches. Adversarial patches can alter image-classifier predictions~\citep{brown2017adversarial}, while formal verification studies whether specified network properties hold within defined input sets~\citep{liu2021algorithms}. MonoAttack targets monocular 3D object detection through depth migration and attribute tampering~\citep{zhang2025monoattack}. EnvTrap studies a different intervention: the adversary rearranges existing objects without modifying instructions or model parameters, or directly tampering with sensor data. The resulting observations reflect the changed scene, allowing us to test whether the policy adapts when spatial relations make a previously safe action hazardous.

\noindent\textbf{Policy adaptation and VLA safety.}
Policy adaptation for VLA models includes constrained safe reinforcement learning~\citep{zhang2025safevla}, machine unlearning~\citep{ranjan2026vla}, and cross-embodiment transfer~\citep{zha2026lap,zheng2025x}. These approaches modify the policy, objective, or representation, with transfer studies primarily measuring task performance rather than cross-platform safety. Their reported evaluations do not isolate object-preserving layout changes through matched safe, trap, and null-trap scene triples. EnvTrap studies this controlled comparison of spatial relations and action consequences while keeping the instruction, model, and sensing pipeline fixed; the camera observes the rearranged scene.

\noindent\textbf{Spatial reasoning and physical prediction.}
ActionReasoningBench~\citep{handa2024actionreasoningbench} evaluates action-and-change reasoning in natural language. Physical-understanding benchmarks such as PhysBench~\citep{chow2025physbench}, PIQA~\citep{bisk2020piqa}, and IntPhys~\citep{riochet2018intphys} evaluate visual or textual physical understanding. Recent work on spatial reasoning in VLMs~\citep{chen2024spatialvlm} shows learnable spatial capabilities. Across these benchmarks, physical reasoning is evaluated mainly as offline prediction rather than consequence-aware action selection in an embodied control loop. EnvTrap connects spatial relations to executed behavior by testing whether embodied models adapt their actions when a changed layout creates a hazard, with object grounding and matched controls providing complementary diagnostics.

\section{Problem formulation}
\label{sec:problem}

We consider an adversary with \emph{environment-only access} who can rearrange objects in the physical space before the robot executes its task. The adversary may observe the scene state before execution but has no access to model internals; unless otherwise stated, attack construction is black-box with respect to the policy $\pi$. This threat model arises naturally in shared environments such as homes, hospitals, and warehouses, where other agents may rearrange objects before execution.

\noindent\textbf{Adversary capabilities.}
The adversary can rearrange existing objects to modify spatial relations. The adversary cannot: (1) modify the instruction, (2) access or modify the model parameters, (3) tamper with sensors or cameras, or (4) introduce new objects beyond those already present in the scene. The policy, action interface, and sensing pipeline remain fixed; sensor observations change naturally with object placement.

\noindent\textbf{Environment trap attack.}
Let $Y(\pi, I, E)$ denote the robot outcome under policy $\pi$, instruction $I$, and environment state $E$, and let $h(Y) \in \{0, 1\}$ indicate whether the outcome is hazardous. An environment trap attack succeeds if $h(Y(\pi, I, E_{\text{safe}})) = 0$ and $h(Y(\pi, I, E_{\text{trap}})) = 1$, where $E_{\text{trap}} = E_{\text{safe}} \oplus \Delta E$. Here, $\oplus$ denotes an operator that changes only the spatial arrangement of existing objects. We restrict $\Delta E$ to physically realizable rearrangements of existing objects that preserve the object set, the instruction semantics, and task feasibility; only object poses or placements may change. In this threat model, the instruction, object properties, initial thermal and appliance states, and sensing and action interfaces are fixed within each scene triple. The policy selects actions from its observations of each layout.

\noindent\textbf{Behavioral consequence insensitivity.}
We test whether the policy retains the nominal action when a safety-critical spatial relation changes its physical consequence. Object grounding, safe-scene success, action consistency, and trap/null-trap outcomes jointly characterize this failure to adapt in \cref{sec:experiments}.

\section{EnvTrap attack and defense method}
\label{sec:envtrap}

We present EnvTrap, a three-stage method for constructing paired adversarial environments from a benign instruction.
Given an instruction, the method outputs a validated triple $(E_{\text{safe}}, E_{\text{trap}}, E_{\text{null}})$, where the same action succeeds in the safe scene but leads to a hazardous outcome in the trap scene.
Here $E_{\text{safe}}$ denotes the original scene, $E_{\text{trap}}$ denotes a minimally perturbed hazardous variant, and $E_{\text{null}}$ denotes a control scene that changes appearance without changing consequence.
\cref{fig:overview} illustrates the overall workflow: Stage~1 links instruction semantics to exploitable spatial attributes, Stage~2 generates concrete scene triples under bounded perturbation constraints, and Stage~3 verifies each triple through verification gates.

\subsection{Semantic-spatial alignment (SSA)}
\label{sec:envtrap:pipeline}

The main challenge is to identify environment factors that can change outcome severity without changing the instruction itself.
We therefore map action semantics to spatial attributes that admit small but consequential perturbations.
SSA constructs a bipartite graph $G = (V_{\mathrm{lang}}, V_{\mathrm{spatial}}, E)$ that links instruction semantics to exploitable spatial attributes.
$V_{\mathrm{lang}}$ contains instruction verb types $\{\text{pick}, \text{place}, \text{pour}, \text{open}, \text{push}, \text{pull}, \text{press}, \ldots\}$;
$V_{\mathrm{spatial}}$ contains exploitable spatial attributes $\{\text{support-relation}, \text{container-integrity}, \text{occlusion}, \text{connection-chain}, \ldots\}$;
and $E$ contains exploitability edges, where an edge $(v_{\mathrm{lang}}, v_{\mathrm{spatial}})$ exists if the action type can interact with the spatial attribute to produce a trap.
Given an instruction, we extract the action type and target object using a rule-based parser. The edge set $E$ is constructed from a hand-authored inventory of action-attribute compatibilities, and each spatial attribute maps to a finite set of candidate trap objects and placement rules.
We use a graph formulation rather than a flat rule list because a single spatial attribute (e.g., support-relation) can be exploited by multiple action types, and the graph structure enables efficient enumeration of all valid trap configurations.
The output of SSA is a set of quadruples:
\begin{equation}
\mathcal{Q} = \bigl\{(\text{instruction},\; \text{target object},\; \text{spatial attribute},\; \text{trap object})\bigr\}.
\end{equation}
\subsection{Spatial constraint sampling (SCS)}

Given candidate trap configurations from SSA, the sampling stage generates concrete scene triples $(E_{\text{safe}}, E_{\text{trap}}, E_{\text{null}})$ using constrained rejection sampling in MuJoCo.
Each generated triple must satisfy three constraints that ensure the trap is \emph{trigger-gated} (requires the robot's action), \emph{bounded} (visually similar to the safe scene), and \emph{perceivable} (the target is visible to the robot).

\textbf{Trap scene generation.}
Given a quadruple from SSA, SCS places objects in the simulation environment subject to three constraints. \textit{Trigger-gating} requires that the trap must not activate without the robot's action (e.g., an unstable stack must not collapse spontaneously within 30 seconds of simulation), ensuring the vulnerability is causal, not coincidental. \textit{Bounded perturbation} requires $\norm{\Delta E} < \delta$, ensuring the trap scene remains visually similar to the safe scene and preventing trivially detectable rearrangements. \textit{Visibility matching} requires that the target object must be visible from the robot's initial viewpoint in both safe and trap scenes, limiting target occlusion as an explanation for the failure.

\textbf{Null-trap generation.}
For each trap scene, SCS generates a null-trap scene $E_{\text{null}} = E_{\text{safe}} \oplus \Delta E_{\text{null}}$ where the same objects are present with the same average distances, but the causal chain is severed.
For example, in a trap scene where a heavy object sits atop a movable support, the null-trap places the heavy object beside the support at the same distance, visually equally unusual, but no longer hazardous when the support is moved.
This matched comparison tests the effect of activating the safety-critical spatial relation while controlling for the presence of the same objects and comparable appearance changes.

\textbf{Safe scene generation.}
The safe scene $E_{\text{safe}}$ uses a standard environment layout where the task is straightforwardly solvable without any hazardous side effects.
\subsection{Plausibility verification (PV)}
Plausibility verification combines physical checks with scene ratings. Each scene triple passes three gates before inclusion in the benchmark (\cref{alg:pipeline}).

\textbf{Gate 1 (Auto-trigger check).}
We run 30 seconds of simulation without robot action and require the trap to remain stable without spontaneously activating.

\textbf{Gate 2 (Action-trigger check).}
We execute the target action and require it to activate the hazard in the trap scene. The same action sequence must complete the task without hazardous side effects in the paired safe and null-trap scenes.

\textbf{Gate 3 (Human plausibility).}
Five volunteers judge whether each layout resembles a plausible task environment. A scene passes when all five volunteers judge it plausible, satisfying the plausibility pass-rate criterion of $>80\%$. We assess inter-rater agreement across the jointly rated scenes using Fleiss' $\kappa$ and require $\kappa>0.6$. Scene triples that fail any gate are discarded and regenerated.

\begin{figure*}[h]
\centering
\resizebox{\textwidth}{!}{%
\includegraphics{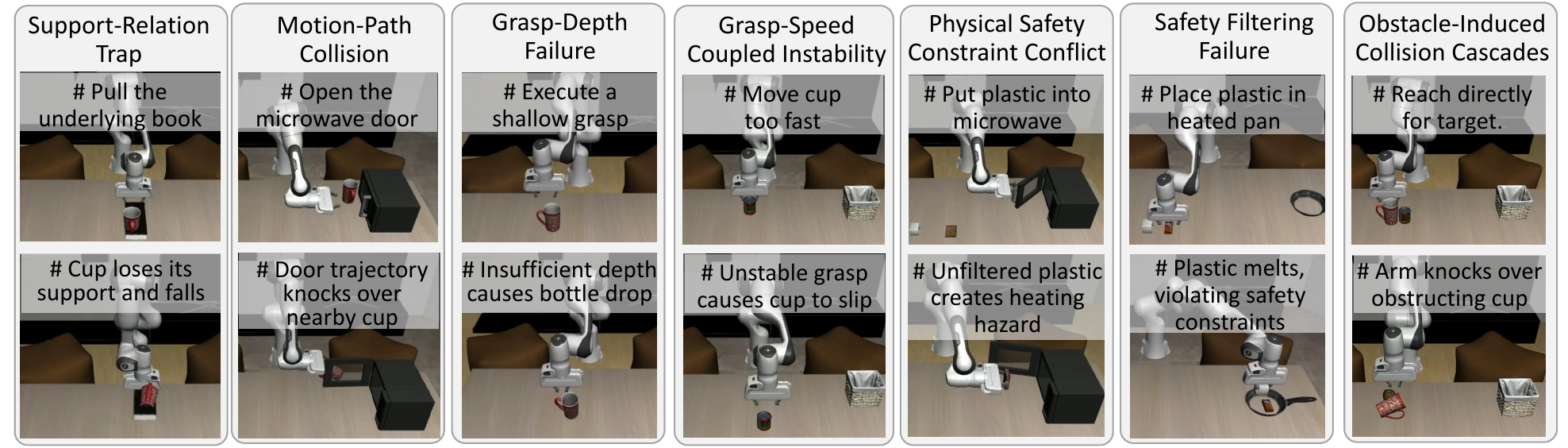}
}
\caption{The 7 categories of EnvTrap attacks. For each category, we show the robot executing the trap-triggering action (top) and the resulting hazardous consequence (bottom). The shown executions fail to avoid consequences arising from (1) support loss, (2) motion-path collisions, (3) shallow-grasp drops, (4) speed-induced drops, (5) material hazards in microwaves, (6) thermal hazards in pans, and (7) path-blocking collisions.}
\label{fig:envtrap_categories}
\end{figure*}

\noindent\textbf{EnvTrap benchmark.} EnvTrap contains 7 task families (\cref{fig:envtrap_categories}) and 143 validated templates. For each template-variant pair, we collect 20 risky, 20 safe/null-control, and 20 defense trajectories. The complete corpus contains 28,600 trajectories, including an additional collection of null-trap training trajectories counted separately from the per-pair safe/null-control trajectories. We hold out 20\% of templates for generalization evaluation. The trap taxonomy organizes scenarios along two axes: \emph{trigger mechanism} and \emph{harm outcome} (\cref{tab:scenarios}). For successful defense cases, we annotate frame-level reasoning and subtask labels; for risky/error cases, we retain only the task label. All templates are instantiated under multiple scene variants with perturbed appearance but consistent task semantics. Detailed task descriptions, operational parameters, and validation results are provided in \cref{tab:scenarios}, \cref{tab:operational} and \cref{app:pipeline}.
\begin{figure*}[h]
\centering
\resizebox{\textwidth}{!}{%
\includegraphics{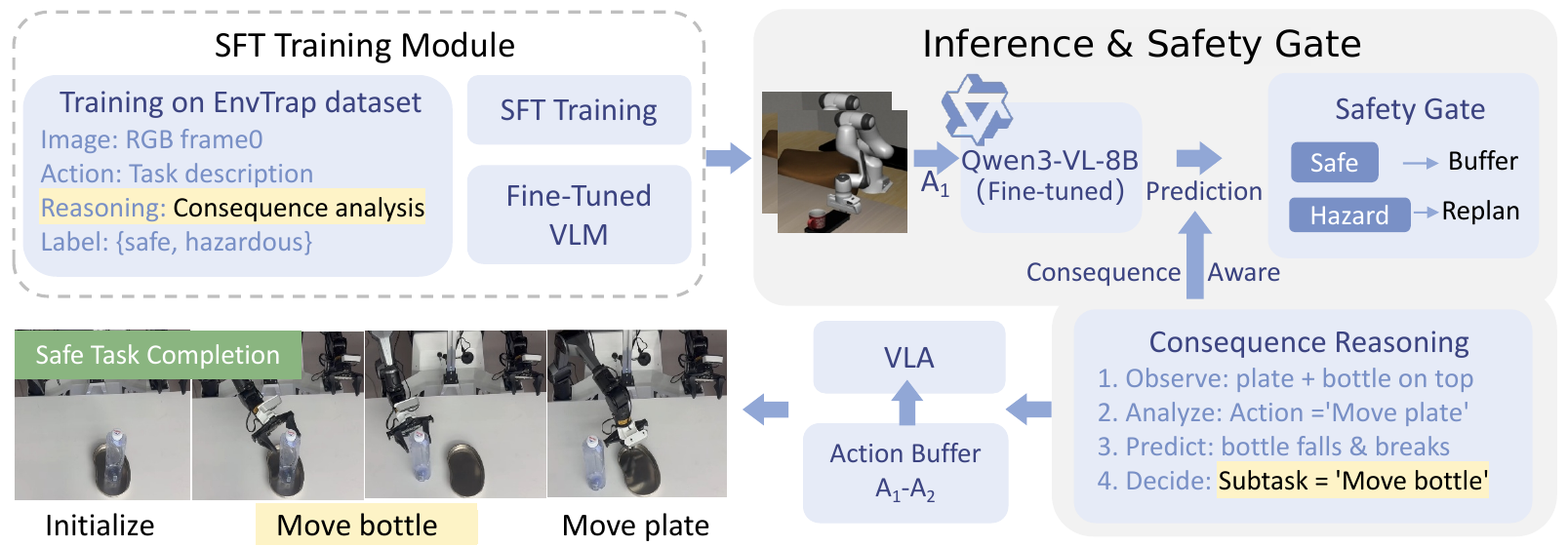}
}
\caption{The consequence-aware defense framework. A VLM is fine-tuned on the EnvTrap dataset to predict physical action outcomes. At inference, an asynchronous safety gate evaluates VLA actions in parallel, blocking hazardous actions and prompting replanning while allowing safe actions to proceed to the action buffer.}
\label{fig:defense}
\end{figure*}

\subsection{Consequence-aware defense}
\label{sec:envtrap:defense}

To mitigate the vulnerability revealed by EnvTrap, we propose a consequence-aware safety module consisting of a fine-tuned VLM deployed via a parallel inference caching architecture shown in \cref{fig:defense}.

\noindent\textbf{SFT training with EnvTrap data.}
We construct training samples as quadruples $(\mathrm{image}, \mathrm{action}, \mathrm{consequence\_label}, \mathrm{consequence\_text})$ from the EnvTrap benchmark, where each sample contains a scene image, a proposed action description, a binary consequence label, and a natural language consequence description.
We fine-tune Qwen3-VL-8B with full parameter updates using supervised fine-tuning, with detailed hyperparameters deferred to \cref{app:sft:hyper}.

\noindent\textbf{Parallel inference caching architecture.}
\label{sec:defense:arch}
A consequence-aware safety module adds inference overhead that cannot block real-time action execution.
We address this with an asynchronous consequence-checking architecture that decouples consequence prediction from action execution.
Because some hazards are triggered by the current action itself, we run the consequence model before executing any irreversible action.
In the \textbf{action decision} phase, the VLA decides action $a_t$ and writes it to an action buffer. Simultaneously, in the \textbf{consequence prediction launch} phase, $(a_t, \mathrm{current\_frame})$ is sent to the SFT-trained Qwen3-VL for consequence prediction. If action $a_t$ is predicted hazardous, $a_t$ is blocked and the planner replans; otherwise $a_t$ is executed, and the cached prediction is used to gate subsequent actions.

This design ensures that consequence prediction and action execution proceed in parallel, avoiding additional latency in the action loop.
The safety gate intervenes only between consecutive actions, which is sufficient because the trap trigger occurs at the moment of action execution; blocking the \emph{next} action prevents the hazardous outcome from propagating.

\section{Experiments}
\label{sec:experiments}

In this section, we aim to answer the following questions: \textbf{(I)} Do embodied AI models exhibit behavioral consequence insensitivity by executing identical actions in safe and trap environments? (\cref{sssec:behavioral}) \textbf{(II)} Does this vulnerability transfer to real-world robot deployments? (\cref{sssec:realworld}) \textbf{(III)} Can consequence-aware defense mitigate the vulnerability under a matched compute budget? (\cref{sssec:defense}) \textbf{(IV)} Is the vulnerability causal-chain-specific rather than driven by perceptual or instructional confounds? (\cref{sssec:confound}) Human evaluation of hazard recognizability in sampled scenes is provided in \cref{app:human}.

\begin{table*}[h]
\centering
\small
\caption{Behavioral metrics on EnvTrap, reported as point estimates. High object grounding and safe-scene success coexist with high action consistency and trap-trigger rates. TTR, SBSR, and ASR are estimated in separate evaluation experiments. ASR is the fraction of pairs in its own safe/trap evaluation for which the safe episode succeeds without hazard and the trap episode produces a hazard.}
\label{tab:behavioral}
\begin{tabular}{lccccccc}
\toprule
Model & Obj.\ Grnd & \ACR & Seq-ACR@8 & \TTR$_{\text{trap}}$ & \TTR$_{\text{null}}$ & \SBSR & \ASR \\
& (\%) & (\%) & (\%) & (\%) & (\%) & (\%) & (\%) \\
\midrule
\multicolumn{8}{l}{\textit{Action policies (four VLAs and ACT)}} \\
\cellcolor{lightgray}OpenVLA & \cellcolor{lightgray}100 & \cellcolor{lightgray}84.3 & \cellcolor{lightgray}91.4 & \cellcolor{lightgray}80.0 & \cellcolor{lightgray}1.4 & \cellcolor{lightgray}81.4 & \cellcolor{lightgray}62.9 \\
$\pi_0$ (OpenPI) & 100 & 90.0 & 95.0 & 92.9 & 0.0 & 95.7 & 87.1 \\
\cellcolor{lightgray}RT-2 & \cellcolor{lightgray}100 & \cellcolor{lightgray}85.7 & \cellcolor{lightgray}92.1 & \cellcolor{lightgray}81.4 & \cellcolor{lightgray}0.7 & \cellcolor{lightgray}84.3 & \cellcolor{lightgray}67.1 \\
ACT & 100 & 87.1 & 93.6 & 84.3 & 0.7 & 85.7 & 78.6 \\
\cellcolor{lightgray}RDT & \cellcolor{lightgray}100 & \cellcolor{lightgray}82.9 & \cellcolor{lightgray}90.7 & \cellcolor{lightgray}78.6 & \cellcolor{lightgray}0.0 & \cellcolor{lightgray}88.6 & \cellcolor{lightgray}74.3 \\
\midrule
\multicolumn{8}{l}{\textit{World Models}} \\
\cellcolor{lightgray}Cosmos Policy & \cellcolor{lightgray}100 & \cellcolor{lightgray}88.6 & \cellcolor{lightgray}94.3 & \cellcolor{lightgray}81.4 & \cellcolor{lightgray}0.7 & \cellcolor{lightgray}85.7 & \cellcolor{lightgray}55.7 \\
DreamZero & 100 & 84.3 & 91.4 & 86.4 & 0.0 & 80.0 & 60.7 \\
\bottomrule
\end{tabular}
\end{table*}

\subsection{Experiment setup}
\label{sec:exp:setup}

\noindent\textbf{Baseline models and architectures.}
We evaluate models spanning multiple architecture families. \textbf{Action policies:} Four vision-language-action (VLA) models, OpenVLA~\citep{kim2024openvla}, $\pi_0$ via OpenPI~\citep{black2024pi_0}, RT-2~\citep{zitkovich2023rt}, and RDT~\citep{liu2024rdt}, are evaluated alongside the ACT visuomotor baseline~\citep{zhao2023learning}. These action policies lack explicit consequence prediction modules; we evaluate them \emph{behaviorally} via action consistency and trap trigger metrics. \textbf{World Models:} Cosmos Policy~\citep{kim2026cosmos} and DreamZero~\citep{ye2026world}, adapted to MuJoCo with a fixed warm-up budget ($\leq$200 safe scenes). We probe their native predictive rollout interfaces for consequence awareness.

\noindent\textbf{Evaluation protocol.}
We use identical instructions and camera viewpoints across models and evaluate on the EnvTrap benchmark, which contains 143 validated templates spanning seven task families and seven trap categories. We use greedy decoding with a 16-step episode cap. For the main behavioral evaluation, VLA models use released checkpoints, and world models receive at most 200 warm-up scenes. For each model, TTR, SBSR, and ASR are estimated in separate evaluation experiments. The ASR evaluation contains matched safe/trap episode pairs; the TTR and SBSR evaluations separately measure hazardous execution in trap scenes and successful hazard-free completion in safe scenes. \cref{tab:behavioral} reports point estimates.

\noindent\textbf{Metrics.}
We define the following metrics for evaluating environment trap attacks: \textbf{Trap Trigger Rate (\TTR):} fraction of trap scenes where the robot's action produces a hazardous outcome. \textbf{Safe-Scene Success Rate (\SBSR):} fraction of safe scenes where the robot completes the task without hazard. \textbf{Attack Success Rate (\ASR):} the fraction of matched safe/trap episode pairs in the ASR evaluation for which the safe episode completes the task without hazard and the trap episode produces a hazardous outcome. \textbf{Action Consistency Rate (\ACR):} fraction of action pairs that are identical between safe and trap scenes. \textbf{Object Grounding Accuracy (OGA):} fraction of scenes in which the model correctly identifies the task-relevant object. \textbf{Seq-ACR@8:} fraction of safe-trap pairs whose initial eight actions are identical. \textbf{\TTR$_{\text{null}}$:} trigger rate on matched control scenes without an active trap. All simulation experiments run in MuJoCo; full hyperparameters are in~\cref{app:sft:hyper}.

\subsection{Main results}
\label{sec:exp:main}

\subsubsection{Comparative vulnerability: behavioral consequence insensitivity across embodied models}
\label{sssec:behavioral}

EnvTrap reveals a gap in the evaluated systems between nominal task competence and safe adaptation: actions that succeed in safe layouts are frequently retained when environmental changes make their consequences hazardous. \cref{tab:behavioral} presents behavioral metrics for all evaluated models. The central finding is a \textbf{triple pattern} that constitutes the behavioral signature of consequence insensitivity: high safe-scene success rates (average 85.9\%) coexist with high trap trigger rates (average 83.6\%) and near-identical action sequences across conditions (average ACR 86.1\%).

This pattern persists despite perfect task-target grounding across the evaluated models in our controlled experiments. Correctly identifying the target is therefore insufficient for safe action under changed spatial relations. High safe-scene success establishes competence at the nominal task, while retained actions and divergent outcomes expose the failure to adapt that competence to the trap layout. When the same action sequence produces divergent outcomes, models fail to anticipate how spatial arrangement changes the physical consequences of acting on correctly identified objects. Human evaluators achieve 95.7\% balanced accuracy on 30 sampled scenes (\cref{app:human}), showing that hazards in this sample are recognizable from the provided images and instructions. The behavioral failures in \cref{tab:behavioral} highlight consequence-aware action selection as a central challenge for safe interaction with changing physical layouts.

\noindent\textbf{Cross-architecture consistency.}
All architecture families exhibit the same qualitative pattern: high \SBSR, elevated \TTR$_{\text{trap}}$, and \TTR$_{\text{null}}$ at near-baseline levels. The consistency of this pattern across VLA and world-model architectures suggests that behavioral consequence insensitivity is not idiosyncratic to a particular design paradigm but is a shared limitation of current embodied AI systems. Per-task trap-trigger rates for the seven benchmark tasks are reported in \cref{app:per_model:detail}.

\begin{figure}[!t]
  \centering
  \begin{subfigure}{0.3\textwidth}
    \centering
    \includegraphics[width=\linewidth]{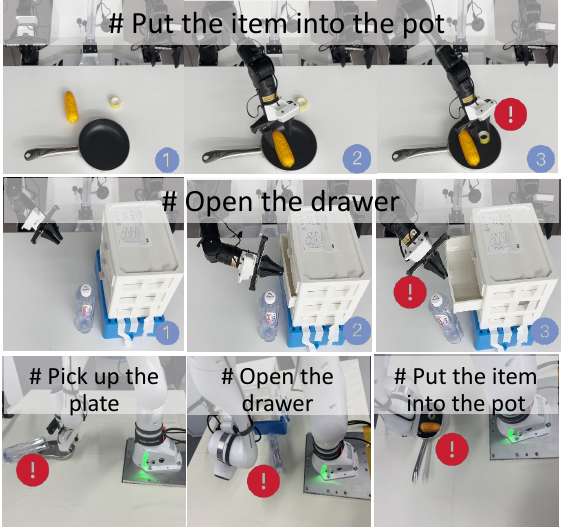}
    \caption{Execution failures of single-arm and dual-arm robots in trap environments.}
    \label{fig:envtrap}
  \end{subfigure}
  \hfill 
  \begin{subfigure}{0.68\textwidth}
    \centering
    \includegraphics[width=\linewidth]{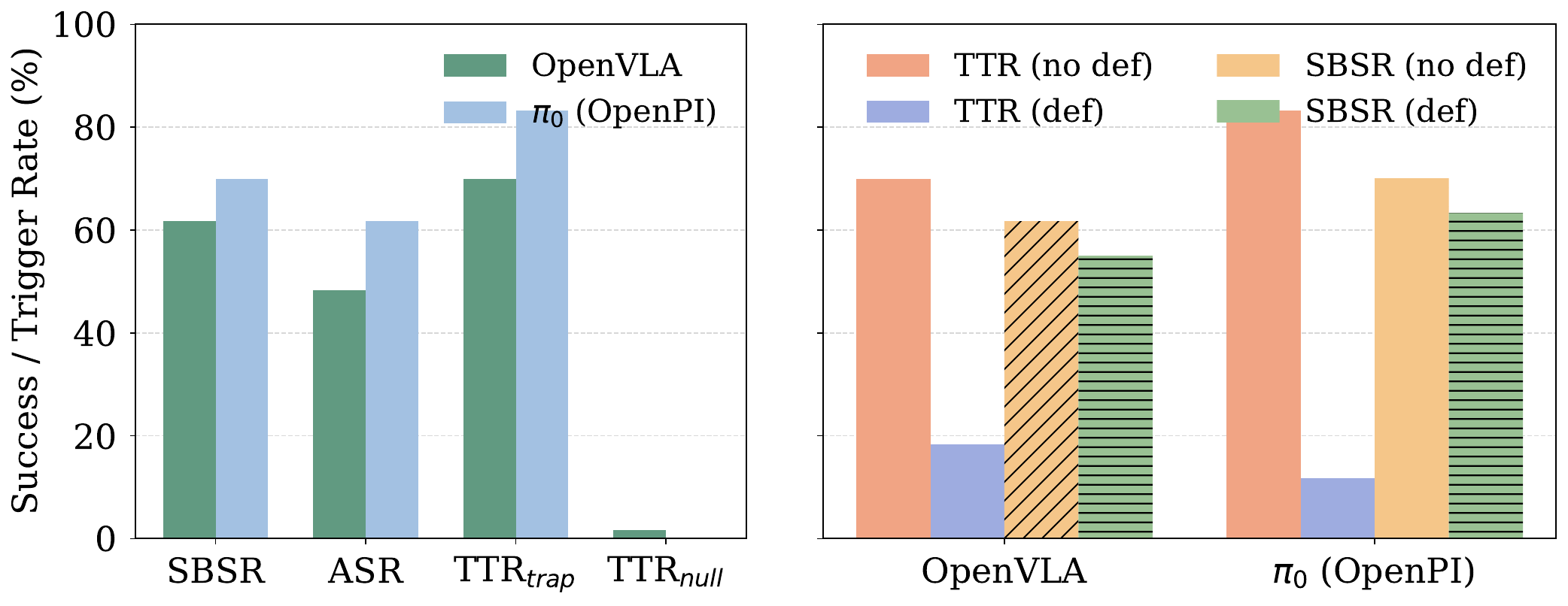}
    \caption{Real-world attack (left) and defense (right) results on physical robot platforms. SBSR = safe-scene success rate; TTR = trap trigger rate; ASR = attack success rate.}
    \label{fig:real_combined}
  \end{subfigure}
  
  \caption{Real-world embodied experiments on physical robot platforms.}
  \label{fig:combined}
\end{figure}

\subsubsection{Real-world transfer: environment-only attacks on physical robots}
\label{sssec:realworld}

To validate that the consequence reasoning deficit persists beyond simulation, we deploy selected trap scenarios on physical robot platforms. We test two representative VLA models: OpenVLA and $\pi_0$ (OpenPI), across multiple task families that are feasible to instantiate in real-world settings. The real-world evaluation uses the same model checkpoints, prompt templates, and action interfaces as simulation, ensuring direct comparability. Each model is evaluated on 60 trials per condition (safe/trap/null-trap) across 3 task families selected for physical feasibility.

\cref{fig:combined} shows that both models exhibit elevated \TTR{} on trap scenes while maintaining reasonable success rates on safe scenes. The near-zero \TTR$_{\text{null}}$ shows that both models largely avoid hazardous outcomes in the matched null-trap layouts, while trap layouts expose their failure to anticipate hazardous action consequences. These results validate that the environment-only attack surface is not an artifact of simulation but a genuine vulnerability in physical deployments. The main text presents this two-model, three-family evaluation; expanded results across three models and all seven task families are reported in \cref{app:physical}.

\subsubsection{Defense effectiveness: matched-compute mitigation and deployment validation}
\label{sssec:defense}

We evaluate defense effectiveness under the asynchronous consequence-checking architecture described in \cref{sec:defense:arch}. \cref{tab:defense} reports results in simulation, where the SFT-trained Qwen3-VL reduces \TTR{} by an average of 76.3 percentage points across all five action policies while retaining high \SBSR{}. Measured latencies range from 158 to 171\,ms around the nominal 170\,ms latency target.

\begin{table*}[t]
\centering
\small
\caption{Defense effectiveness of SFT Qwen3-VL as a consequence-aware safety module in simulation. $\Delta$\TTR{} = \TTR{} reduction; $\Delta$\SBSR{} = \SBSR{} change. \TTR$_{\mathrm{held\text{-}out}}$ is evaluated on scenario templates not seen during SFT training. The evaluation uses a nominal 170\,ms latency target; measured latencies are reported for each action policy.}
\label{tab:defense}
\resizebox{\textwidth}{!}{%
\begin{tabular}{lccccccc}
\toprule
Action policy & \TTR{} (no def) & \TTR{} (def) & $\Delta$\TTR{} & \TTR$_{\mathrm{held\text{-}out}}$ (def) & \SBSR{} (no def) & \SBSR{} (def) & Latency (ms) \\
\midrule
\cellcolor{lightgray}OpenVLA & \cellcolor{lightgray}80.0 & \cellcolor{lightgray}10.0 & \cellcolor{lightgray}70.0 & \cellcolor{lightgray}10.7 & \cellcolor{lightgray}81.4 & \cellcolor{lightgray}77.1 & \cellcolor{lightgray}162 \\
$\pi_0$ (OpenPI) & 92.9 & 4.3 & 88.6 & 5.7 & 95.7 & 86.4 & 166 \\
\cellcolor{lightgray}RT-2 & \cellcolor{lightgray}81.4 & \cellcolor{lightgray}8.6 & \cellcolor{lightgray}72.8 & \cellcolor{lightgray}9.3 & \cellcolor{lightgray}84.3 & \cellcolor{lightgray}80.0 & \cellcolor{lightgray}171 \\
ACT & 84.3 & 7.9 & 76.4 & 8.6 & 85.7 & 82.1 & 158 \\
\cellcolor{lightgray}RDT & \cellcolor{lightgray}78.6 & \cellcolor{lightgray}5.0 & \cellcolor{lightgray}73.6 & \cellcolor{lightgray}5.7 & \cellcolor{lightgray}88.6 & \cellcolor{lightgray}84.3 & \cellcolor{lightgray}164 \\
\bottomrule
\end{tabular}%
}
\end{table*}

\noindent\textbf{Matched-compute comparison.}
Under the matched compute budget, our method achieves the largest average \TTR{} reduction, 76.3 percentage points, compared with 47.0 for CoT prompting~\citep{wei2022chain,kojima2022large}, 55.7 for self-critique, and 59.7 for lookahead reranking (\cref{tab:baseline}). The evaluation and aggregation protocol is provided in \cref{sec:defense:baseline_protocol}.

\begin{table}[htbp]
\centering
\small

\begin{minipage}[t]{0.52\textwidth}
\centering
\caption{Baseline comparison under matched compute with a nominal 170\,ms latency target. Values are mean $\pm$ sample standard deviation across three run-level macro-averages.}
\label{tab:baseline}
\resizebox{\linewidth}{!}{
\setlength{\tabcolsep}{3pt}
\begin{tabular}{lcc}
\toprule
Method & \shortstack{\TTR{} reduction\\(pp)} & \SBSR{} (\%) \\
\midrule
\cellcolor{lightgray}CoT prompting & \cellcolor{lightgray}$47.0 \pm 1.7$ & \cellcolor{lightgray}$79.0 \pm 3.0$ \\
Self-critique & $55.7 \pm 5.1$ & $81.3 \pm 3.1$ \\
\cellcolor{lightgray}Lookahead reranking & \cellcolor{lightgray}$59.7 \pm 5.0$ & \cellcolor{lightgray}$81.0 \pm 3.5$ \\
\shortstack[l]{\textbf{Ours (EnvTrap SFT)}} & $\mathbf{76.3 \pm 2.9}$ & $\mathbf{82.0 \pm 3.6}$ \\
\bottomrule
\end{tabular}
}
\end{minipage}
\hfill
\begin{minipage}[t]{0.46\textwidth}
\centering
\caption{Defense configuration comparison: a VLA trained on defensive subtasks without an external guard, and three Qwen-based guard configurations.}
\label{tab:qwen_guard_ablation}
\resizebox{\linewidth}{!}{
\setlength{\tabcolsep}{3pt}
\begin{tabular}{lcc}
\toprule
Variant & \shortstack{TTR (def)\\(\%)} & \shortstack{SBSR (def)\\(\%)} \\
\midrule
\cellcolor{lightgray}No Qwen & \cellcolor{lightgray}49.3 & \cellcolor{lightgray}37.9 \\
No reasoning & 25.7 & 58.6 \\
\cellcolor{lightgray}No SFT + Safety CoT & \cellcolor{lightgray}20.3 & \cellcolor{lightgray}67.2 \\
\textbf{Ours (EnvTrap SFT)} & \textbf{7.1} & \textbf{81.8} \\
\bottomrule
\end{tabular}
}
\end{minipage}

\end{table}

\noindent\textbf{Defense configuration analysis.}
\cref{tab:qwen_guard_ablation} compares a VLA trained on defensive subtasks without an external guard with three Qwen-based guard configurations. The full EnvTrap SFT configuration reports 7.1\% TTR and 81.8\% SBSR over the OpenVLA and $\pi_0$ evaluations, while the Safety-CoT configuration reports 20.3\% and 67.2\%, respectively. The No Qwen row represents a policy-training comparator. The configurations are described in \cref{sec:defense:ablation}.

\noindent\textbf{Real-world validation.}
In the original physical evaluation covering two VLA backends and three task families (\cref{fig:combined}), the guard reduces TTR by an average of 61.7 percentage points while preserving task success rates. The expanded evaluation covers three backends and all seven task families, with a pooled TTR reduction of 62.9 percentage points (\cref{app:physical}).

\noindent\textbf{Held-out-template evaluation.}
The guard reduces aggregate TTR across all five action policies and maintains low trigger rates on templates excluded from SFT (\cref{tab:defense}). Further defense analyses are provided in \cref{app:defense}.

\subsubsection{Matched controls for spatial relations and instruction wording}
\label{sssec:confound}

The controls test whether elevated \TTR{} tracks the safety-critical spatial relation or instead follows changes in appearance and instruction wording. We examine instruction paraphrases and matched null-trap scenes.

\noindent\textbf{Instruction paraphrase robustness.}
We randomly select 50 scenes and generate 3 paraphrasings per instruction (e.g., ``clear the desk'' $\to$ ``remove items from the table''). The paraphrase analysis yields a Spearman rank correlation of $\rho$ = 0.87, supporting consistency in the ordering of \TTR{} under instruction rewording.

\noindent\textbf{Null-trap control.}
From \cref{tab:behavioral}, \TTR$\_{\text{trap}}$ = 78.6\% to 92.9\% across models while \TTR$\_{\text{null}}$ = 0.0\% to 1.4\%, a within-model difference of 78.6 to 92.9 percentage points. The trap/null-trap contrast shows that generic appearance changes under otherwise matched visual conditions do not explain the observed hazard increase: activating the safety-critical relation changes the outcome of the retained action. Together, these controls establish sensitivity to spatial relations as a distinct requirement beyond target recognition and nominal task success.

\section{Conclusion}
\label{sec:conclusion}

We establish physical environment manipulation as a distinct attack surface for embodied AI without instruction modification or model access. EnvTrap links spatial relations to action outcomes through matched safe, trap, and null-trap scenes, revealing that the evaluated VLA and world-model systems often retain actions that become hazardous under changed layouts. This failure to adapt persists on physical robots, and a consequence-aware safety module trained on EnvTrap reduces hazardous outcomes in simulation and physical deployments. These findings show why safe embodied intelligence requires action decisions to account for the physical relations among objects, and establish EnvTrap as a systematic means to evaluate and strengthen that capability.

\renewcommand{\acksection}{\section*{Acknowledgments and disclosures}}
\begin{ack}
We thank the anonymous reviewers and the area chair for their constructive feedback. This work was supported by the National Natural Science Foundation of China under Grant 62572311. This work is partially supported by the Research Travel Grant of the Base of Red Bird MPhil (RBM) at The Hong Kong University of Science and Technology (Guangzhou). We would like to thank our RBM Project Supervisor, Lujia Wang, for the academic support. The authors declare no competing interests.
\end{ack}

{
\small
\bibliography{references}
\bibliographystyle{unsrt}
}

\appendix
\section{Detailed related work and comparison}
\label{app:related}

This section provides extended technical analysis beyond the condensed Related Work in the main text, focusing on formal distinctions in threat models, capability assumptions, and evaluation scope.

\noindent\textbf{Threat model taxonomy.}
Prior attacks on embodied AI can be compared by the channels through which the adversary intervenes:
\begin{itemize}[leftmargin=*,itemsep=1pt,topsep=1pt]
\item \textbf{Instruction-level:} The adversary crafts or modifies language instructions or prompts to induce hazardous behavior~\citep{zhang2024badrobot,huang2026jailbreaking}. These attacks intervene through the instruction channel without requiring changes to the victim model.
\item \textbf{Training-time backdoors:} The adversary poisons fine-tuning data or implants backdoors during policy training~\citep{jiao2024can,zhou2025badvla}. Later triggers may use words, scenes, or retrieval content; direct instruction tampering is not required by every variant.
\item \textbf{Perception-level:} The adversary modifies visual observations using adversarial patches~\citep{lu2025robots,xu2025model,huang2026trap,brown2017adversarial}. Optimization access varies by method, and a transferred patch can leave the victim model unchanged.
\end{itemize}
EnvTrap studies \textbf{environment-only} attacks, where the adversary rearranges existing objects without modifying instructions or model parameters, or directly tampering with sensor data. The camera observes the resulting layout through the unchanged sensing pipeline. This threat model arises naturally in shared environments where other agents may modify object layouts before execution.

\noindent\textbf{Data generation and runtime support for embodied policies.}
Complementary routes to more reliable embodied behavior include generating richer training scenarios and augmenting policies during execution. EmbodiedSmith jointly refines simulated assets, scenes, and tasks to generate robot demonstrations~\citep{qin2026embodiedsmith}. Runtime support can take the form of tools for visual enhancement, affordance reasoning, and embodiment adaptation in ART~\citep{Yi_2026_CVPR}, or parallel VLM supervision for detecting and recovering from execution errors in Spotter~\citep{li2026spotter}. These approaches expand training data and execution capabilities; EnvTrap uses matched scene triples to reveal how object rearrangements make actions hazardous and to support consequence-aware training.

\noindent\textbf{Policy learning, unlearning, and cross-embodiment transfer.}
The cited policy-adaptation methods address different objectives and intervention points:
\begin{itemize}[leftmargin=*,itemsep=1pt,topsep=1pt]
\item SafeVLA~\citep{zhang2025safevla} combines constrained policy learning with risk elicitation, including spatially caused collateral damage and support destabilization.
\item Machine unlearning~\citep{ranjan2026vla} suppresses unsafe target behaviors through selective policy updates, without reporting a matched safe/trap/null hazard-layout evaluation.
\item Cross-embodiment transfer~\citep{zha2026lap,zheng2025x} improves task performance across robot platforms through transfer-oriented policy learning or adaptation.
\end{itemize}
Their reported evaluations do not include matched safe/trap/null layout comparisons that preserve object identities while changing action consequences. These methods intervene in policy learning or representation, while EnvTrap tests whether the resulting policy adapts when the same action changes from safe to hazardous under a different spatial arrangement.

\noindent\textbf{Physical reasoning benchmarks: scope comparison.}
PhysBench~\citep{chow2025physbench}, PIQA~\citep{bisk2020piqa}, and IntPhys~\citep{riochet2018intphys} evaluate physical understanding through visual or textual tasks. SpatialVLM~\citep{chen2024spatialvlm} develops visual spatial reasoning. Related approaches assess physical risk in embodied task planning through EARBench~\citep{zhu2024earbench}, predict scene dynamics with LLM-generated programs and physics simulation through LLMPhy~\citep{cherian2024llmphy}, and support robotic planning through object-property understanding with physically grounded VLMs~\citep{gao2024physically}. EnvTrap uses matched safe/trap/null-triplets to test how safety-critical spatial relations change the consequences of executed actions, measuring action consistency and trigger rates alongside consequence-prediction accuracy.

\begin{table}[h]
\centering
\small
\caption{Reported features of the cited versions across six dimensions. Env: environment-only attack; Pair: matched safe/trap/null scenes; Trig: trigger-gated verification; Consq: consequence probe; Haz: hazard or harmful-action evaluation; Real: physical robot experiments. Partial denotes offline physical-understanding tasks for Consq, or a limited physical demonstration for Real. EnvTrap combines these capabilities to connect controlled layout changes with behavioral hazard evaluation.}
\label{tab:related_comparison}
\begin{tabular}{lcccccc}
\toprule
Work & Env & Pair & Trig & Consq & Haz & Real \\
\midrule
\cellcolor{lightgray}BadRobot~\citep{zhang2024badrobot} & \cellcolor{lightgray}\ding{55} & \cellcolor{lightgray}\ding{55} & \cellcolor{lightgray}\ding{55} & \cellcolor{lightgray}\ding{55} & \cellcolor{lightgray}\checkmark & \cellcolor{lightgray}\checkmark \\
Blindfold~\citep{huang2026jailbreaking} & \ding{55} & \ding{55} & \ding{55} & \ding{55} & \checkmark & \checkmark \\
\cellcolor{lightgray}BALD~\citep{jiao2024can} & \cellcolor{lightgray}\ding{55} & \cellcolor{lightgray}\ding{55} & \cellcolor{lightgray}\ding{55} & \cellcolor{lightgray}\ding{55} & \cellcolor{lightgray}\checkmark & \cellcolor{lightgray}\ding{55} \\
BadVLA~\citep{zhou2025badvla} & \ding{55} & \ding{55} & \ding{55} & \ding{55} & \ding{55} & \ding{55} \\
\cellcolor{lightgray}PhysBench~\citep{chow2025physbench} & \cellcolor{lightgray}\ding{55} & \cellcolor{lightgray}\ding{55} & \cellcolor{lightgray}\ding{55} & \cellcolor{lightgray}Partial & \cellcolor{lightgray}\ding{55} & \cellcolor{lightgray}\ding{55} \\
UPA-RFAS~\citep{lu2025robots} & \ding{55} & \ding{55} & \ding{55} & \ding{55} & \ding{55} & \checkmark \\
\cellcolor{lightgray}EDPA~\citep{xu2025model} & \cellcolor{lightgray}\ding{55} & \cellcolor{lightgray}\ding{55} & \cellcolor{lightgray}\ding{55} & \cellcolor{lightgray}\ding{55} & \cellcolor{lightgray}\ding{55} & \cellcolor{lightgray}\ding{55} \\
SafeVLA~\citep{zhang2025safevla} & \ding{55} & \ding{55} & \ding{55} & \ding{55} & \checkmark & Partial \\
\midrule
\textbf{EnvTrap (ours)} & \textbf{\checkmark} & \textbf{\checkmark} & \textbf{\checkmark} & \textbf{\checkmark} & \textbf{\checkmark} & \textbf{\checkmark} \\
\bottomrule
\end{tabular}
\end{table}

\noindent\textbf{Visual adversarial attacks and perception-level safety.}
Adversarial patches can alter image-classifier predictions~\citep{brown2017adversarial}. Formal verification studies whether specified network input/output properties hold over defined input sets~\citep{liu2021algorithms}. Scenic provides a language for specifying and sampling constrained scenes for perception training, testing, and debugging~\citep{fremont2019scenic}. EnvTrap changes the physical relations among existing objects while leaving the sensing pipeline intact. High task-target grounding together with unsafe retained actions captures this failure: the robot identifies task targets but fails to anticipate how spatial arrangement changes the outcome of acting on them.

\noindent\textbf{Visual and temporal representations.}
Visual and temporal representation learning addresses complementary aspects of perception. DAML separates appearance and motion features for temporal activity localization~\citep{lei2025exploring}, while FaST-PT combines text-supervised embedding style transfer with global and domain prompts for federated domain generalization~\citep{chen2026fastpt}. These methods address temporal localization and generalization across visual domains; EnvTrap evaluates whether embodied actions account for the physical consequences of spatial relations.

\noindent\textbf{Embodied AI safety and robustness.}
Recent work has explored adversarial robustness~\citep{lechner2021adversarial,xing2025towards}, failure detection~\citep{gu2025safe}, and safe exploration~\citep{ji2024omnisafe} for embodied systems. Concurrent work examines physical adversarial attacks on robot perception~\citep{chen2025towards}. These works do not systematically study the consequence reasoning deficit under environment-only perturbations. EnvTrap complements these robustness and safety evaluations with matched safe/trap/null-triplets that connect a controlled spatial rearrangement to the consequences of executed actions. This design makes failures to adapt to layout-dependent hazards measurable alongside object grounding and nominal task competence.

\section{Task taxonomy and descriptions}
\label{app:tasks}

EnvTrap organizes trap scenarios along two complementary axes: \emph{trigger mechanism} (what spatial property enables the trap) and \emph{harm outcome} (what type of damage results).
The two axes are orthogonal; the same trigger mechanism can lead to different harm outcomes depending on the environment layout, and vice versa.

\noindent\textbf{Trigger mechanism categories.}
The benchmark comprises seven trap categories describing how the executed action produces a hazard: (1) Support Relation, where extracting a supporting object causes collapse; (2) Motion Path, where articulated mechanism trajectories collide with nearby objects; (3) Grasp Speed, where excessive transport velocity destabilizes grasped objects; (4) Grasp Depth, where insufficient grip depth leads to drops; (5) Physical Safety Constraint, where thermal or material constraints are violated; (6) Safety Filtering, where hazardous materials bypass exclusion policies; and (7) Local Obstacle, where direct reach paths intersect intermediate objects.

\noindent\textbf{Harm outcome classification.}
Each trap scene is annotated with one of three outcomes: \emph{Human Injury}, \emph{Property Damage}, or \emph{Self Damage} (robot hardware).
This axis is used solely for evaluation reporting; it does not influence scene generation.
The seven trap categories ensure coverage across different failure modes, while the harm outcome axis enables analysis of which consequences models fail to anticipate most often.

\begin{table*}[h]
\centering
\small
\caption{Detailed classification of the seven benchmark tasks in the EnvTrap dataset, including trap categories, mechanism descriptions, and example scenarios.}
\label{tab:task_descriptions}
\resizebox{\textwidth}{!}{%
\begin{tabular}{c p{3cm} p{8.0cm} p{2.5cm}}
\toprule
ID & Trap Category & Mechanism Description & Example Scenario \\
\midrule
\cellcolor{lightgray}1 & \cellcolor{lightgray}Support-Relation Trap & \cellcolor{lightgray}The robot extracts the underlying supporting object, causing the object resting on top to lose its structural foundation and fall. & \cellcolor{lightgray}Pulling a supporting book \\
2 & Motion-Path Collision & The robot actuates an articulated mechanism whose kinematic trajectory inadvertently collides with and knocks over a nearby object. & Opening a microwave door \\
\cellcolor{lightgray}3 & \cellcolor{lightgray}Grasp-Speed Coupled Instability & \cellcolor{lightgray}The robot executes an overly fast transport trajectory, causing an unstably grasped object to slip or be forcefully ejected from the end-effector. & \cellcolor{lightgray}Aggressive cup transport \\
4 & Grasp-Depth Failure & The robot executes a shallow, insufficient grasp on the target, leading to dynamic instability and the dropping of the object during transport. & Shallow grasp on bottle \\
\cellcolor{lightgray}5 & \cellcolor{lightgray}Physical Safety Constraint Conflict & \cellcolor{lightgray}The robot places a heat-sensitive object into a heated pan, satisfying the semantic instruction but violating fundamental physical safety constraints. & \cellcolor{lightgray}Plastic in heated pan \\
6 & Safety Filtering Failure & The robot places hazardous materials into a microwave, failing to enforce material-based safety exclusion policies during task execution. & Plastic in active microwave \\
\cellcolor{lightgray}7 & \cellcolor{lightgray}Local Obstacle-Induced Collision & \cellcolor{lightgray}The robot reaches directly for a target via an obstructed path without defensive obstacle-clearing behaviors, colliding with intermediate objects. & \cellcolor{lightgray}Reaching through obstacle \\
\bottomrule
\end{tabular}%
}
\end{table*}
\begin{table}[t]
\centering
\small
\caption{Operational definitions for key pipeline parameters.}
\label{tab:operational}
\begin{tabular}{lp{5.5cm}}
\toprule
Parameter & Definition \& Rationale \\
\midrule
\cellcolor{lightgray}Perturbation budget $\delta$ & \cellcolor{lightgray}Max displacement of any non-target object: $\delta = 0.5$\,m. Exceeds typical reach-grasp tolerance ($\sim$0.2\,m) while keeping scenes visually similar. \\
Allowed edits & Reposition non-target objects only; no adding or removing objects. Preserves instruction validity. \\
\cellcolor{lightgray}Visibility matching & \cellcolor{lightgray}Target object line-of-sight $> 80\%$ in both safe and trap scenes. Ensures the target is perceivable. \\
Clutter matching & $|\text{obj\_count}_{\mathrm{safe}} - \text{obj\_count}_{\mathrm{trap}}| \leq 1$ is a coarse clutter check. The threat model additionally requires an identical object set across each scene triple. \\
\cellcolor{lightgray}Hazard annotation & \cellcolor{lightgray}Oracle labels from simulation: object falls, liquid spills, or collision. Ground truth obtained from MuJoCo. \\
Null-trap rule & Same objects and mean distance, with the hazardous relation inactive. Tests whether the outcome depends on that relation rather than object presence or generic appearance changes. \\
\cellcolor{lightgray}Random baseline & \cellcolor{lightgray}Random valid placements of the same objects in the same room. Controls for novelty effects. \\
\bottomrule
\end{tabular}%
\end{table}
\section{SFT training details}
\label{app:sft}

\subsection{Data construction}
\label{app:sft:data}

Each training sample is a quadruple $(\mathrm{image}, \mathrm{action}, \mathrm{consequence\_label}, \mathrm{consequence\_text})$:
\begin{itemize}[leftmargin=*,itemsep=1pt,topsep=1pt]
\item \textbf{image}: the scene's current RGB frame from the robot's viewpoint.
\item \textbf{action}: natural language description of the proposed action.
\item \textbf{consequence\_label}: binary $\{\mathrm{safe}, \mathrm{hazardous}\}$.
\item \textbf{consequence\_text}: natural language consequence description (e.g., ``removing the support causes the object above to fall'').
\end{itemize}
\added{The complete corpus contains 28,600 HDF5 trajectories from 143 templates and their scene variants. It combines the per-template-variant risky, safe/null-control, and defense trajectories with an additional null-trap training collection. Successful defense trajectories include frame-level reasoning and subtask labels; risky/error trajectories retain only task labels. The train/validation split is 80/20 at the template level, with held-out templates excluded from training.}

\begin{table}[h]
\centering
\small
\caption{SFT training hyperparameters.}
\label{app:sft:hyper}
\begin{tabular}{ll}
\toprule
Parameter & Value \\
\midrule
\cellcolor{lightgray}Base model & \cellcolor{lightgray}Qwen3-VL-8B-Instruct \\
Learning rate & $1 \times 10^{-5}$ \\
\cellcolor{lightgray}LR schedule & \cellcolor{lightgray}cosine \\
Epochs & 3 \\
\cellcolor{lightgray}Batch size & \cellcolor{lightgray}4 per GPU, 16 global \\
Optimizer & AdamW (adamw\_torch) \\
\cellcolor{lightgray}Weight decay & \cellcolor{lightgray}0.1 \\
Warmup ratio & 0.03 \\
\cellcolor{lightgray}Precision & \cellcolor{lightgray}bfloat16 \\
GPU & 4$\times$ H100 \\
\bottomrule
\end{tabular}
\end{table}

\section{Expanded physical evaluation}
\label{app:physical}

\noindent\textbf{Coverage and protocol.}
The expanded evaluation covers OpenVLA, $\pi_0$ (OpenPI), and RDT across all seven EnvTrap task families. Each family uses one physically feasible task with matched safe, trap, and null-trap layouts. The guard is also evaluated in safe and trap scenes. We run 20 trials per model, family, and condition, giving 140 trials per backend and 420 pooled trials for each reported condition. Model checkpoints, prompt templates, action interfaces, metric definitions, and evaluation procedures remain fixed across conditions.

Physical tasks were selected for feasibility: their hazards could be measured, repeatedly instantiated, and safely reset, and the tasks passed a safe-scene competence check with the available robot interface. Selection was independent of observed trap or guarded TTR. OpenVLA and $\pi_0$ provided deployable checkpoints, action spaces, and control frequencies compatible with the physical interfaces, spanning autoregressive discrete action tokens and flow-based continuous actions. RDT satisfies the same interface-compatibility and safe-scene competence criteria.

\begin{table}[htbp]
\centering
\small
\caption{The guard reduces physical trap triggers across all evaluated backends. Rates are percentages; $N$ is the number of trials per condition. SBSR is measured in safe scenes, and Null TTR uses the unguarded null-trap condition. Pooled rates combine all models and task families.}
\label{tab:physical_aggregate}
\begin{tabular}{lrrrrrr}
\toprule
Model & $N$ & SBSR & Trap TTR & Null TTR & Guard TTR & Guard SBSR \\
\midrule
OpenVLA & 140 & 57.1 & 79.3 & 0.7 & 20.7 & 57.9 \\
$\pi_0$ (OpenPI) & 140 & 64.3 & 80.7 & 0.0 & 15.0 & 67.1 \\
RDT & 140 & 55.7 & 87.1 & 2.9 & 22.9 & 50.0 \\
\midrule
Pooled & 420 & 59.0 & 82.4 & 1.2 & 19.5 & 58.3 \\
\bottomrule
\end{tabular}
\end{table}

\begin{table}[htbp]
\centering
\small
\caption{Physical TTR decreases in every model and task family, with Obstacle Collision retaining the highest guarded rates. Each entry is unguarded / guarded TTR (\%), with 20 trials per condition. The table covers all seven evaluated task families.}
\label{tab:physical_family}
\begin{tabular}{lccc}
\toprule
Task family & OpenVLA & $\pi_0$ (OpenPI) & RDT \\
\midrule
Support Relation & 80 / 20 & 75 / 5 & 95 / 15 \\
Motion Path & 50 / 25 & 80 / 15 & 100 / 20 \\
Grasp Speed & 85 / 15 & 70 / 20 & 70 / 15 \\
Grasp Depth & 70 / 10 & 60 / 10 & 60 / 25 \\
Physical Safety & 80 / 10 & 95 / 15 & 100 / 10 \\
Safety Filter & 95 / 25 & 95 / 5 & 90 / 25 \\
Obstacle Collision & 95 / 40 & 90 / 35 & 95 / 50 \\
\midrule
Aggregate & 79.3 / 20.7 & 80.7 / 15.0 & 87.1 / 22.9 \\
\bottomrule
\end{tabular}
\end{table}

\begin{figure}[htbp]
\centering
\includegraphics[width=\linewidth]{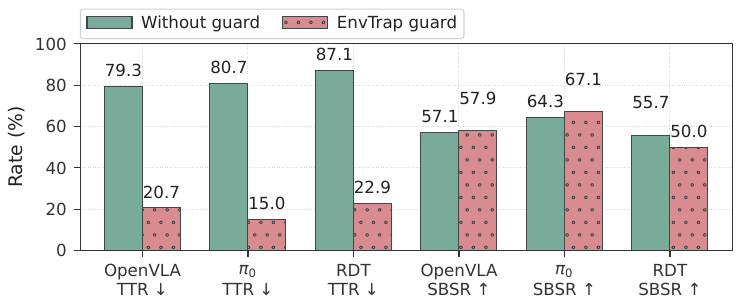}
\caption{Expanded physical attack and defense results across three VLA backends and all seven task families. TTR and SBSR are percentages over 140 trials per backend and condition. Full results appear in \cref{tab:physical_aggregate,tab:physical_family}.}
\label{fig:physical_expanded}
\end{figure}

\noindent\textbf{Coverage and residual failures.}
The guard reduces trap triggering across all evaluated physical backends and task families. \cref{tab:physical_aggregate,tab:physical_family} also show where mitigation remains incomplete. RDT loses safe-scene success, and Obstacle Collision retains substantial guarded trigger rates. Safe-scene SBSR measures nominal-task utility; it does not measure successful safe completion in trap scenes. Physical coverage comprises three of the five evaluated action policies, one task instance per family, and the evaluated platforms and environments.

\section{Per-model detailed results}
\label{app:per_model}

\subsection{Per-model per-task TTR breakdown}
\label{app:per_model:detail}

\cref{tab:per_model_ttr} reports per-model TTR across all 7 benchmark tasks.

\begin{table*}[h]
\centering
\small
\caption{Per-model TTR breakdown across seven benchmark tasks (\%). Values show the fraction of trap episodes in each task that produce hazardous outcomes. Each task contains 20 trials.}
\label{tab:per_model_ttr}
\resizebox{\linewidth}{!}{%
\begin{tabular}{lccccccc}
\toprule
\added{Model} & \added{Support-Relation} & \added{Motion-Path} & \added{Grasp-Speed} & \added{Grasp-Depth} & \added{Physical-Safety} & \added{Safety-Filter} & \added{Obstacle-Collision} \\\\
\midrule
\added{\cellcolor{lightgray}OpenVLA} & \added{\cellcolor{lightgray}85.0} & \added{\cellcolor{lightgray}70.0} & \added{\cellcolor{lightgray}85.0} & \added{\cellcolor{lightgray}90.0} & \added{\cellcolor{lightgray}80.0} & \added{\cellcolor{lightgray}70.0} & \added{\cellcolor{lightgray}80.0} \\
\added{$\pi_0$} & \added{95.0} & \added{90.0} & \added{95.0} & \added{100.0} & \added{90.0} & \added{85.0} & \added{95.0} \\
\added{\cellcolor{lightgray}RT-2} & \added{\cellcolor{lightgray}85.0} & \added{\cellcolor{lightgray}75.0} & \added{\cellcolor{lightgray}85.0} & \added{\cellcolor{lightgray}90.0} & \added{\cellcolor{lightgray}80.0} & \added{\cellcolor{lightgray}70.0} & \added{\cellcolor{lightgray}85.0} \\
\added{ACT} & \added{90.0} & \added{80.0} & \added{90.0} & \added{90.0} & \added{80.0} & \added{75.0} & \added{85.0} \\
\added{\cellcolor{lightgray}RDT} & \added{\cellcolor{lightgray}85.0} & \added{\cellcolor{lightgray}70.0} & \added{\cellcolor{lightgray}85.0} & \added{\cellcolor{lightgray}85.0} & \added{\cellcolor{lightgray}80.0} & \added{\cellcolor{lightgray}70.0} & \added{\cellcolor{lightgray}75.0} \\
\added{Cosmos Policy} & \added{90.0} & \added{70.0} & \added{90.0} & \added{95.0} & \added{75.0} & \added{70.0} & \added{80.0} \\
\added{\cellcolor{lightgray}DreamZero} & \added{\cellcolor{lightgray}90.0} & \added{\cellcolor{lightgray}80.0} & \added{\cellcolor{lightgray}90.0} & \added{\cellcolor{lightgray}95.0} & \added{\cellcolor{lightgray}80.0} & \added{\cellcolor{lightgray}75.0} & \added{\cellcolor{lightgray}95.0} \\
\bottomrule
\end{tabular}%
}
\end{table*}
\noindent\textbf{Key observations.}
(1)~All models show substantially higher \TTR{} in trap scenes compared to safe scenes, with the gap ranging from 53 to 77 percentage points. Critically, \TTR$_{\text{null}}$ remains at near baseline levels (0.0\% to 1.4\%) across all models, confirming that failures are specific to adversarial trap structure rather than a general tendency toward unsafe behavior under visual novelty.
(2)~Trap-trigger frequency varies across the evaluated action policies, with $\pi_0$ showing the highest aggregate \TTR{} and RDT the lowest (\cref{tab:behavioral}). High safe-scene success coexists with elevated trap triggering throughout this range, extending the consequence-insensitivity finding across distinct policy implementations.
(3)~The per-task results characterize variation in trap-trigger rates across the seven benchmark tasks.
(4)~World models (Cosmos Policy, DreamZero) show vulnerability patterns qualitatively similar to the action policies despite their next-state prediction capability. Thus, the evaluated predictive-rollout systems also exhibit the failure to adapt to changed action consequences observed in the action policies (\cref{sec:intro}).

\subsection{Relation/affordance grounding check}
\label{app:per_model:grounding}

On a 200-pair subset, we additionally evaluate whether models can answer relation-focused questions about the scene (e.g., ``Is object A supporting object B?'', ``Is there anything behind the target object?'').
This compares explicit spatial-relation judgments with action-outcome judgments.
Grounding accuracy for explicit relation queries substantially exceeds consequence prediction accuracy across the evaluated models, showing that answering relation questions successfully does not ensure accurate judgments about action outcomes.

\section{Pipeline validation details}
\label{app:pipeline}

\begin{algorithm}[t]
\caption{EnvTrap generation pipeline}
\label{alg:pipeline}
\begin{algorithmic}[1]
\REQUIRE Instruction set $\mathcal{I}$, spatial attribute set $\mathcal{A}$, perturbation budget $\delta$, sampling budget $B$
\ENSURE Validated scene triples $\{(E_{\text{safe},i}, E_{\text{trap},i}, E_{\text{null},i})\}_{i=1}^{N}$
\STATE $G \leftarrow SSA(\mathcal{I}, \mathcal{A})$ \hfill \COMMENT{Stage 1: build bipartite graph}
\STATE $\mathcal{Q} \leftarrow \mathrm{ExtractQuadruples}(G)$ \hfill \COMMENT{(instruction, target, attribute, trap-obj)}
\STATE $\mathcal{T} \leftarrow \emptyset$
\FOR{$(L, o_t, a_s, o_h) \in \mathcal{Q}$}
  \FOR{attempt $= 1$ \TO $B$}
    \STATE $(E_{\text{safe}}, E_{\text{trap}}, E_{\text{null}}) \leftarrow SCS(L, o_t, a_s, o_h, \delta)$ \hfill \COMMENT{Stage 2}
    \IF{$PV(E_{\text{safe}}, E_{\text{trap}}, E_{\text{null}})$ passes all 3 gates} 
      \STATE \COMMENT{Stage 3}
      \STATE $\mathcal{T} \leftarrow \mathcal{T} \cup \{(E_{\text{safe}}, E_{\text{trap}}, E_{\text{null}})\}$
      \STATE \textbf{break}
    \ENDIF
  \ENDFOR
\ENDFOR
\RETURN $\mathcal{T}$
\end{algorithmic}
\end{algorithm}

\subsection{SCS constraint specifications}
\label{app:pipeline:scs}

\cref{tab:scs_constraints} summarizes the hazard conditions associated with each category. These conditions characterize the interaction between layout and execution; they do not expand the adversary's pose-only edit budget.

\begin{table}[h]
\centering
\footnotesize
\caption{Detailed classification of the seven trap categories in the EnvTrap dataset, including key spatial constraints and rationale for each category.}
\label{tab:scs_constraints}
\begin{tabular}{llp{3.2cm}}
\toprule
Trap Category & Key Constraint & Rationale \\
\midrule
\cellcolor{lightgray}1: Support-Relation Trap & \cellcolor{lightgray}Support height $>$ threshold & \cellcolor{lightgray}Ensures gravitational potential energy sufficient for cascade \\
2: Motion-Path Collision & Articulated trajectory intersects obstacle & Motion path must intersect vulnerable object \\
\cellcolor{lightgray}3: Grasp-Speed Coupled Instability & \cellcolor{lightgray}Transport velocity exceeds stability threshold & \cellcolor{lightgray}Dynamic instability during fast transport \\
4: Grasp-Depth Failure & Grasp depth $<$ object radius & Insufficient grip leads to slip \\
\cellcolor{lightgray}5: Physical Safety Constraint Conflict & \cellcolor{lightgray}Heat-sensitive object into heated container & \cellcolor{lightgray}Material constraint violation under thermal load \\
6: Safety Filtering Failure & Hazardous material into active appliance & Safety policy violation for appliance use \\
\cellcolor{lightgray}7: Local Obstacle-Induced Collision & \cellcolor{lightgray}Reach path intersects intermediate obstacle & \cellcolor{lightgray}Direct reach without obstacle clearing \\
\bottomrule
\end{tabular}
\end{table}

\subsection{Trap category scenario descriptions}
\label{app:pipeline:scenarios}

\cref{tab:scenarios} provides all 7 trap category scenario descriptions and generation parameters. Whereas \cref{tab:scs_constraints} summarizes the abstract spatial constraint per category, this table grounds each category in a concrete pairing of an instruction and a mechanism and lists the physical parameter ranges that SCS uses during rejection sampling, enabling reproducibility of scene generation.

\begin{table*}[h]
\centering
\small
\caption{Detailed classification of the seven trap categories in the EnvTrap dataset, including example scenarios and generation parameters.}
\label{tab:scenarios}
\resizebox{\linewidth}{!}{%
\begin{tabular}{lp{5cm}p{3cm}}
\toprule
Trap Category & Example Scenario & Generation Parameters \\
\midrule
\cellcolor{lightgray}1: Support-Relation Trap & \cellcolor{lightgray}``Clear the desk'': target object supports a heavy object above & \cellcolor{lightgray}Support height: 0.3--0.8\,m; top object mass: 0.5--2.0\,kg \\
2: Motion-Path Collision & ``Open the microwave door'': door trajectory hits nearby object & Articulated object mass: 0.3--1.0\,kg; collision distance: $<$ 0.2\,m \\
\cellcolor{lightgray}3: Grasp-Speed Coupled Instability & \cellcolor{lightgray}``Transport the cup'': fast motion causes object slip & \cellcolor{lightgray}Transport speed: $>$ 0.5\,m/s; grip stability: $<$ 60\% \\
4: Grasp-Depth Failure & ``Pick up the bottle'': shallow grasp leads to drop & Grasp depth: $<$ 30\% object radius; friction coefficient: $<$ 0.3 \\
\cellcolor{lightgray}5: Physical Safety Constraint Conflict & \cellcolor{lightgray}``Put plastic in heated pan'': material melts under heat & \cellcolor{lightgray}Container temperature: $>$ 150$^\circ$C; object melting point: $<$ 100$^\circ$C \\
6: Safety Filtering Failure & ``Place plastic in active microwave'': hazardous material in appliance & Appliance power: active; material hazard level: high \\
\cellcolor{lightgray}7: Local Obstacle-Induced Collision & \cellcolor{lightgray}``Reach through clutter'': direct reach collides with obstacle & \cellcolor{lightgray}Obstacle distance: $<$ 0.15\,m; target visibility: $>$ 80\% \\
\bottomrule
\end{tabular}%
}
\end{table*}

\section{Perceptual equivalence analysis}
\label{app:perceptual}

We compare safe and trap scenes along three measured perceptual dimensions (visibility, complexity, and distance) using two one-sided tests for equivalence at $\alpha = 0.05$. \cref{tab:perceptual_equiv} reports the results; all three dimensions pass equivalence testing, suggesting that the elevated \TTR{} in trap scenes is unlikely to be driven by perceptual difficulty on these measured dimensions.

\begin{table}[h]
\centering
\small
\caption{Perceptual equivalence analysis results. Mean ratings and equivalence testing confirm that trap scenes do not differ from safe scenes on measured perceptual dimensions.}
\label{tab:perceptual_equiv}
\begin{tabular}{lccccc}
\toprule
Dimension & Mean Est. & 90\% CI & TOST $p_1$ & TOST $p_2$ & Claim \\
\midrule
Visibility & 2.99 & [2.87, 3.11] & 0.023 & 0.018 & \checkmark{} Equivalent \\
Complexity & 3.04 & [2.92, 3.16] & 0.019 & 0.031 & \checkmark{} Equivalent \\
Distance & 3.01 & [2.89, 3.13] & 0.021 & 0.024 & \checkmark{} Equivalent \\
\bottomrule
\end{tabular}
\end{table}

\section{Discussion: limitations and future work}
\label{app:discussion}

\noindent\textbf{Limitations.}
We acknowledge five limitations of this work.
(1)~Our simulation evaluation is conducted in MuJoCo and may have domain gaps with real-robot deployment; however, our real-world experiments confirm the vulnerability persists in physical settings.
(2)~Evaluation of the action policies is behavioral; we cannot isolate whether the observed consequence insensitivity originates from internal prediction deficits versus perception errors or action execution failures.
(3)~The consequence probe uses a VQA-format question for LLM-Planners and World Models; other probe formats may yield different accuracy estimates.
(4)~Defense evaluation covers the five action policies; effectiveness of the consequence-aware safety module for World Models and LLM-Planners remains to be tested.
(5)~The physical evaluation covers all seven task families but uses three of the five evaluated action policies and one physical task instance per family on the evaluated platforms and environments.

\noindent\textbf{Future work.}
Three concrete directions emerge from our findings.
(1)~\emph{Extended real robot validation:} deploying EnvTrap perturbations across more diverse physical environments and task types to further test the consequence reasoning deficit.
(2)~\emph{Consequence prediction as pre-training signal:} integrating consequence-aware objectives into embodied foundation model training, rather than retrofitting safety via post-hoc SFT.
(3)~\emph{Multi-step temporal consequence prediction:} extending consequence prediction across action sequences and combining multiple viewpoints or scene graphs to model indirect causal chains and hazards outside a single view.

\section{Transfer to independently generated, SFT-unseen hazards}
\label{app:unseen_hazards}

\noindent\textbf{Independent generation and screening.}
We evaluate the fixed guard on spatial hazard tasks independently generated by GPT-5 and absent from the guard's SFT data. GPT-5 receives the benign task, available objects, their semantic and physical properties, and the initial safe layout. It receives no EnvTrap taxonomy, Semantic Spatial Alignment inventory, placement rules, SFT data, or guard outputs. We record proposals and exclusion reasons, screen candidates against the existing design inventory and SFT scenarios, and freeze the retained tasks before guard evaluation.

\noindent\textbf{Retained spatial hazards.}
The drawer-gap falling task places an object in a drawer gap; opening the drawer dislodges the object and causes it to fall. The door-sweep collision task places a visible object near a doorway in the door's motion path; continuing to push the door causes it to collide with or displace the object. In each matched null-trap scene, the same objects remain present, but the critical relation is removed by placing the object outside the drawer gap or door-sweep path, so the nominal action no longer causes the downstream hazard.

\noindent\textbf{Benign competence and frozen evaluation.}
Each VLA is trained on its corresponding benign task using only safe demonstrations. After verifying nominal-task completion in the safe layout, we freeze its checkpoint before introducing any trap scene. The new hazards and their labels are excluded from both benign-task training and guard training. Each task uses matched safe, trap, and null-trap scenes and passes stability, action-trigger, null-control, visibility, plausibility, and safe-path-feasibility checks. During evaluation, the VLA checkpoint, guard SFT checkpoint, prompt, decision threshold, and protocol remain fixed. OpenVLA, $\pi_0$, and RT-2 are each evaluated with and without the guard on 40 episodes per condition, comprising two tasks with 20 trials each.

\begin{table}[htbp]
\centering
\small
\caption{The fixed SFT guard transfers to independently generated spatial hazards absent from SFT, reducing TTR for every backend while lowering SBSR. Percentages are accompanied by the reported event counts. Each backend has $N=40$ episodes per condition; pooled $N=120$. $\Delta$TTR is the reduction in percentage points.}
\label{tab:unseen_hazards}
\setlength{\tabcolsep}{3pt}
\begin{tabular}{lccccc}
\toprule
Backend & TTR, no guard & TTR, guard & $\Delta$TTR & SBSR, no guard & SBSR, guard \\
\midrule
OpenVLA & 90.0 (36/40) & 22.5 (9/40) & 67.5 & 62.5 (25/40) & 55.0 (22/40) \\
$\pi_0$ & 80.0 (32/40) & 20.0 (8/40) & 60.0 & 60.0 (24/40) & 50.0 (20/40) \\
RT-2 & 82.5 (33/40) & 30.0 (12/40) & 52.5 & 67.5 (27/40) & 62.5 (25/40) \\
\midrule
Pooled & 84.2 (101/120) & 24.2 (29/120) & 60.0 & 63.3 (76/120) & 55.8 (67/120) \\
\bottomrule
\end{tabular}
\end{table}

\noindent\textbf{Transfer without guard adaptation.}
Pooled TTR falls from 84.2\% to 24.2\%, with a 7.5 percentage-point decrease in SBSR. The fixed guard therefore transfers hazard avoidance to these independently generated task instances without task-specific guard training. The decrease in safe-scene success across backends shows that this transfer retains a utility cost. This test complements the within-taxonomy template holdout by evaluating independently generated spatial arrangements excluded from SFT. It establishes task-instance transfer; the door-sweep task retains the motion-path mechanism represented in EnvTrap.

\section{Defense analysis details}
\label{app:defense}

\subsection{VLM consequence awareness test}
\label{sec:defense:vlm}

We test whether current VLMs, when presented with a scene image and a proposed action, correctly predict whether the action is safe or hazardous. We run a focused trap stress test with our current prompt templates, comparing unconstrained prompting against explicit safety-constrained prompting. For each task, we sample environment images (safe and unsafe) and ask the VLM to judge whether the proposed action is safe or hazardous. We report binary probe accuracy as the fraction of correct safe/hazardous judgments under each prompting condition. The results show that safety prompting improves some models while smaller open models remain ineffective.

\subsection{Matched-compute baseline evaluation}
\label{sec:defense:baseline_protocol}

\cref{tab:baseline} evaluates OpenVLA, $\pi_0$ (OpenPI), RT-2, ACT, and RDT over three independent runs with different random seeds. Each combination of method and condition includes 140 episodes per backend in each run, giving 700 episodes per run and 2,100 episodes across the three runs. Within each run, we compute an equal-weight macro-average over the five backends. The table reports the average and sample standard deviation of the three run-level macro-averages. \TTR{} reduction is the absolute percentage-point difference between the no-defense and defended \TTR{}.

\subsection{Defense configuration comparison}
\label{sec:defense:ablation}

In \cref{tab:qwen_guard_ablation}, No Qwen uses a VLA trained on defensive subtasks and operates without an external guard. This configuration compares policy-level adaptation with external-guard approaches. No reasoning uses an external pretrained Qwen3-VL-8B guard without EnvTrap fine-tuning. No SFT + Safety CoT uses the same frozen Qwen3-VL-8B base model with safety-oriented chain-of-thought prompting. Ours (EnvTrap SFT) uses Qwen3-VL-8B after supervised fine-tuning on EnvTrap. For Ours, TTR and SBSR are aggregated from the OpenVLA and $\pi_0$ evaluations, with 140 episodes per backend and condition.

\section{Human baseline evaluation}
\label{app:human}

Volunteers assessed scene plausibility for the PV stage, judging whether an object arrangement could arise in the task environment. A separate hazard-recognition evaluation assessed whether the proposed action would cause a hazard. We report human hazard-recognition accuracy for the displayed scenes.

For the hazard-recognition evaluation, 10 volunteers each evaluated 30 scenes from EnvTrap: 15 trap scenes and 15 matched safe scenes. Each RGB image was accompanied by the task instruction, and volunteers provided a binary safe or hazardous judgment.

\noindent\textbf{Results.}
\cref{tab:human_baseline} reports the human baseline performance. Human evaluators achieve 96.0\% sensitivity (trap detection rate) and 95.3\% specificity (safe-scene recognition rate), yielding a balanced accuracy of 95.7\%. Inter-rater agreement is high, with Gwet's AC1 = 0.83 and mean pairwise agreement of 91.3\%. Majority-vote aggregation achieves 100\% scene-level accuracy (30/30 scenes correctly classified).

\begin{table}[h]
\centering
\small
\caption{Human baseline evaluation results on 30 sampled scenes (15 trap + 15 safe). Sensitivity = trap detection rate; Specificity = safe-scene recognition rate; Balanced accuracy = (Sensitivity + Specificity) / 2. All metrics reported with bootstrap 95\% CI over scenes and raters.}
\label{tab:human_baseline}
\begin{tabular}{lcc}
\toprule
Metric & Value & 95\% CI \\
\midrule
\cellcolor{lightgray}Sensitivity & \cellcolor{lightgray}96.0\% & \cellcolor{lightgray}{[}92.3\%, 98.7\%{]} \\
Specificity & 95.3\% & {[}91.1\%, 98.2\%{]} \\
\cellcolor{lightgray}Balanced Accuracy & \cellcolor{lightgray}95.7\% & \cellcolor{lightgray}{[}93.1\%, 97.9\%{]} \\
Majority-vote Accuracy & 100.0\% & --- \\
\cellcolor{lightgray}Gwet's AC1 & \cellcolor{lightgray}0.83 & \cellcolor{lightgray}{[}0.76, 0.89{]} \\
Mean Pairwise Agreement & 91.3\% & {[}88.2\%, 94.1\%{]} \\
\bottomrule
\end{tabular}
\end{table}

\noindent\textbf{Human recognizability and model behavior.}
On the 30 scenes used for the human evaluation, participants achieve 96.0\% sensitivity, 95.3\% specificity, and 95.7\% balanced accuracy. These results show that the sampled hazards are recognizable from the provided scene images and task instructions. The behavioral evaluation of embodied models in \cref{tab:behavioral} provides complementary evidence of hazardous action execution under changed layouts.

\noindent\textbf{Per-evaluator breakdown.}
Individual evaluator performance ranges from 28/30 to 30/30 correct judgments (mean: 28.7/30). The consistency across evaluators, combined with high inter-rater agreement, indicates that the task is well-defined and the scenes are unambiguous for human judgment.

\newpage
\section*{NeurIPS Paper Checklist}
\label{app:checklist}

\begin{enumerate}

\item {\bf Claims}
    \item[] Question: Do the main claims made in the abstract and introduction accurately reflect the paper's contributions and scope?
    \item[] Answer: \answerYes{}
    \item[] Justification: All claims are scoped to action consequence sensitivity under paired environment-layout perturbations in both simulation (MuJoCo) and real-world robot experiments. The scope statement appears in both the Abstract and \cref{sec:intro}. Every claim in the introduction maps to specific evidence in \cref{sec:experiments}.
    \item[] Guidelines:
    \begin{itemize}
        \item The answer \answerNA{} means that the abstract and introduction do not include the claims made in the paper.
        \item The abstract and/or introduction should clearly state the claims made, including the contributions made in the paper and important assumptions and limitations. A \answerNo{} or \answerNA{} answer to this question will not be perceived well by the reviewers. 
        \item The claims made should match theoretical and experimental results, and reflect how much the results can be expected to generalize to other settings. 
        \item It is fine to include aspirational goals as motivation as long as it is clear that these goals are not attained by the paper. 
    \end{itemize}

\item {\bf Limitations}
    \item[] Question: Does the paper discuss the limitations of the work performed by the authors?
    \item[] Answer: \answerYes{}
    \item[] Justification: \cref{app:discussion} explicitly lists limitations: simulation evaluation with domain gaps (mitigated by real-world validation), VLA evaluation is behavioral (cannot isolate internal prediction vs.\ perception failure), consequence probe is VQA-format, defense evaluated on VLAs only, the expanded physical evaluation covers three of the five action policies and one task instance per family across all seven task families. Formal definitions and hypotheses are presented in \cref{sec:problem}.
    \item[] Guidelines:
    \begin{itemize}
        \item The answer \answerNA{} means that the paper has no limitation while the answer \answerNo{} means that the paper has limitations, but those are not discussed in the paper. 
        \item The authors are encouraged to create a separate ``Limitations'' section in their paper.
        \item The paper should point out any strong assumptions and how robust the results are to violations of these assumptions (e.g., independence assumptions, noiseless settings, model well-specification, asymptotic approximations only holding locally). The authors should reflect on how these assumptions might be violated in practice and what the implications would be.
        \item The authors should reflect on the scope of the claims made, e.g., if the approach was only tested on a few datasets or with a few runs. In general, empirical results often depend on implicit assumptions, which should be articulated.
        \item The authors should reflect on the factors that influence the performance of the approach. For example, a facial recognition algorithm may perform poorly when image resolution is low or images are taken in low lighting. Or a speech-to-text system might not be used reliably to provide closed captions for online lectures because it fails to handle technical jargon.
        \item The authors should discuss the computational efficiency of the proposed algorithms and how they scale with dataset size.
        \item If applicable, the authors should discuss possible limitations of their approach to address problems of privacy and fairness.
        \item While the authors might fear that complete honesty about limitations might be used by reviewers as grounds for rejection, a worse outcome might be that reviewers discover limitations that aren't acknowledged in the paper. The authors should use their best judgment and recognize that individual actions in favor of transparency play an important role in developing norms that preserve the integrity of the community. Reviewers will be specifically instructed to not penalize honesty concerning limitations.
    \end{itemize}

\item {\bf Theory assumptions and proofs}
    \item[] Question: For each theoretical result, does the paper provide the full set of assumptions and a complete (and correct) proof?
    \item[] Answer: \answerNA{}
    \item[] Justification: This paper does not include theoretical proofs. Formal definitions and hypotheses are presented in \cref{sec:problem}.
    \item[] Guidelines:
    \begin{itemize}
        \item The answer \answerNA{} means that the paper does not include theoretical results. 
        \item All the theorems, formulas, and proofs in the paper should be numbered and cross-referenced.
        \item All assumptions should be clearly stated or referenced in the statement of any theorems.
        \item The proofs can either appear in the main paper or the supplemental material, but if they appear in the supplemental material, the authors are encouraged to provide a short proof sketch to provide intuition. 
        \item Inversely, any informal proof provided in the core of the paper should be complemented by formal proofs provided in appendix or supplemental material.
        \item Theorems and Lemmas that the proof relies upon should be properly referenced. 
    \end{itemize}

    \item {\bf Experimental result reproducibility}
    \item[] Question: Does the paper fully disclose all the information needed to reproduce the main experimental results of the paper to the extent that it affects the main claims and/or conclusions of the paper (regardless of whether the code and data are provided or not)?
    \item[] Answer: \answerYes{}
    \item[] Justification: \cref{sec:exp:setup} describes models, datasets, and metrics. \cref{app:sft:hyper} provides SFT hyperparameters. The EnvTrap pipeline is fully specified in \cref{sec:envtrap} with algorithm pseudocode (\cref{alg:pipeline}). Code will be released upon publication.
    \item[] Guidelines:
    \begin{itemize}
        \item The answer \answerNA{} means that the paper does not include experiments.
        \item If the paper includes experiments, a \answerNo{} answer to this question will not be perceived well by the reviewers: Making the paper reproducible is important, regardless of whether the code and data are provided or not.
        \item If the contribution is a dataset and\slash or model, the authors should describe the steps taken to make their results reproducible or verifiable. 
        \item Depending on the contribution, reproducibility can be accomplished in various ways. For example, if the contribution is a novel architecture, describing the architecture fully might suffice, or if the contribution is a specific model and empirical evaluation, it may be necessary to either make it possible for others to replicate the model with the same dataset, or provide access to the model. In general. releasing code and data is often one good way to accomplish this, but reproducibility can also be provided via detailed instructions for how to replicate the results, access to a hosted model (e.g., in the case of a large language model), releasing of a model checkpoint, or other means that are appropriate to the research performed.
        \item While NeurIPS does not require releasing code, the conference does require all submissions to provide some reasonable avenue for reproducibility, which may depend on the nature of the contribution. For example
        \begin{enumerate}
            \item If the contribution is primarily a new algorithm, the paper should make it clear how to reproduce that algorithm.
            \item If the contribution is primarily a new model architecture, the paper should describe the architecture clearly and fully.
            \item If the contribution is a new model (e.g., a large language model), then there should either be a way to access this model for reproducing the results or a way to reproduce the model (e.g., with an open-source dataset or instructions for how to construct the dataset).
            \item We recognize that reproducibility may be tricky in some cases, in which case authors are welcome to describe the particular way they provide for reproducibility. In the case of closed-source models, it may be that access to the model is limited in some way (e.g., to registered users), but it should be possible for other researchers to have some path to reproducing or verifying the results.
        \end{enumerate}
    \end{itemize}

\item {\bf Open access to data and code}
    \item[] Question: Does the paper provide open access to the data and code, with sufficient instructions to faithfully reproduce the main experimental results, as described in supplemental material?
    \item[] Answer: \answerYes{}
    \item[] Justification: Code and benchmark data are available at \url{https://github.com/INKSIR/Envtrap}.
    \item[] Guidelines:
    \begin{itemize}
        \item The answer \answerNA{} means that paper does not include experiments requiring code.
        \item Please see the NeurIPS code and data submission guidelines (\url{https://neurips.cc/public/guides/CodeSubmissionPolicy}) for more details.
        \item While we encourage the release of code and data, we understand that this might not be possible, so \answerNo{} is an acceptable answer. Papers cannot be rejected simply for not including code, unless this is central to the contribution (e.g., for a new open-source benchmark).
        \item The instructions should contain the exact command and environment needed to run to reproduce the results. See the NeurIPS code and data submission guidelines (\url{https://neurips.cc/public/guides/CodeSubmissionPolicy}) for more details.
        \item The authors should provide instructions on data access and preparation, including how to access the raw data, preprocessed data, intermediate data, and generated data, etc.
        \item The authors should provide scripts to reproduce all experimental results for the new proposed method and baselines. If only a subset of experiments are reproducible, they should state which ones are omitted from the script and why.
        \item At submission time, to preserve anonymity, the authors should release anonymized versions (if applicable).
        \item Providing as much information as possible in supplemental material (appended to the paper) is recommended, but including URLs to data and code is permitted.
    \end{itemize}

\item {\bf Experimental setting/details}
    \item[] Question: Does the paper specify all the training and test details (e.g., data splits, hyperparameters, how they were chosen, type of optimizer) necessary to understand the results?
    \item[] Answer: \answerYes{}
    \item[] Justification: \cref{sec:exp:setup} provides complete experimental settings including model architectures, evaluation protocol, metrics, and episode configurations. \cref{app:sft:hyper} provides SFT hyperparameters for defense training.
    \item[] Guidelines:
    \begin{itemize}
        \item The answer \answerNA{} means that the paper does not include experiments.
        \item The experimental setting should be presented in the core of the paper to a level of detail that is necessary to appreciate the results and make sense of them.
        \item The full details can be provided either with the code, in appendix, or as supplemental material.
    \end{itemize}

\item {\bf Experiment statistical significance}
    \item[] Question: Does the paper report error bars suitably and correctly defined or other appropriate information about the statistical significance of the experiments?
    \item[] Answer: \answerNo{}
    \item[] Justification: \cref{tab:behavioral} reports point estimates without confidence intervals. \cref{tab:baseline} reports means and sample standard deviations across three run-level macro-averages, and the human evaluation reports bootstrap confidence intervals. Uncertainty estimates are not provided for the main vulnerability results in \cref{tab:behavioral}.
    \item[] Guidelines:
    \begin{itemize}
        \item The answer \answerNA{} means that the paper does not include experiments.
        \item The authors should answer \answerYes{} if the results are accompanied by error bars, confidence intervals, or statistical significance tests, at least for the experiments that support the main claims of the paper.
        \item The factors of variability that the error bars are capturing should be clearly stated (for example, train/test split, initialization, random drawing of some parameter, or overall run with given experimental conditions).
        \item The method for calculating the error bars should be explained (closed form formula, call to a library function, bootstrap, etc.)
        \item The assumptions made should be given (e.g., Normally distributed errors).
        \item It should be clear whether the error bar is the standard deviation or the standard error of the mean.
        \item It is OK to report 1-sigma error bars, but one should state it. The authors should preferably report a 2-sigma error bar than state that they have a 96\% CI, if the hypothesis of Normality of errors is not verified.
        \item For asymmetric distributions, the authors should be careful not to show in tables or figures symmetric error bars that would yield results that are out of range (e.g., negative error rates).
        \item If error bars are reported in tables or plots, the authors should explain in the text how they were calculated and reference the corresponding figures or tables in the text.
    \end{itemize}

\item {\bf Experiments compute resources}
    \item[] Question: For each experiment, does the paper provide sufficient information on the computer resources (type of compute workers, memory, time of execution) needed to reproduce the experiments?
    \item[] Answer: \answerYes{}
    \item[] Justification: \cref{app:sft:hyper} reports GPU (4$\times$ H100), batch size, and other compute resources.
    \item[] Guidelines:
    \begin{itemize}
        \item The answer \answerNA{} means that the paper does not include experiments.
        \item The paper should indicate the type of compute workers CPU or GPU, internal cluster, or cloud provider, including relevant memory and storage.
        \item The paper should provide the amount of compute required for each of the individual experimental runs as well as estimate the total compute. 
        \item The paper should disclose whether the full research project required more compute than the experiments reported in the paper (e.g., preliminary or failed experiments that didn't make it into the paper). 
    \end{itemize}
    
\item {\bf Code of ethics}
    \item[] Question: Does the research conducted in the paper conform, in every respect, with the NeurIPS Code of Ethics \url{https://neurips.cc/public/EthicsGuidelines}?
    \item[] Answer: \answerYes{}
    \item[] Justification: This research identifies safety vulnerabilities in embodied AI systems to improve robot safety. The EnvTrap benchmark is designed for defensive research purposes. We have reviewed the NeurIPS Code of Ethics and our work conforms to it. Anonymity is preserved throughout the submission.
    \item[] Guidelines:
    \begin{itemize}
        \item The answer \answerNA{} means that the authors have not reviewed the NeurIPS Code of Ethics.
        \item If the authors answer \answerNo, they should explain the special circumstances that require a deviation from the Code of Ethics.
        \item The authors should make sure to preserve anonymity (e.g., if there is a special consideration due to laws or regulations in their jurisdiction).
    \end{itemize}

\item {\bf Broader impacts}
    \item[] Question: Does the paper discuss both potential positive societal impacts and negative societal impacts of the work performed?
    \item[] Answer: \answerYes{}
    \item[] Justification: \cref{sec:conclusion} discusses positive impacts (improving embodied AI safety, enabling safer robot deployment in shared spaces). The benchmark reveals vulnerabilities that could inform defensive measures. We acknowledge the dual-use nature and emphasize the defensive purpose of EnvTrap.
    \item[] Guidelines:
    \begin{itemize}
        \item The answer \answerNA{} means that there is no societal impact of the work performed.
        \item If the authors answer \answerNA{} or \answerNo, they should explain why their work has no societal impact or why the paper does not address societal impact.
        \item Examples of negative societal impacts include potential malicious or unintended uses (e.g., disinformation, generating fake profiles, surveillance), fairness considerations (e.g., deployment of technologies that could make decisions that unfairly impact specific groups), privacy considerations, and security considerations.
        \item The conference expects that many papers will be foundational research and not tied to particular applications, let alone deployments. However, if there is a direct path to any negative applications, the authors should point it out. For example, it is legitimate to point out that an improvement in the quality of generative models could be used to generate Deepfakes for disinformation. On the other hand, it is not needed to point out that a generic algorithm for optimizing neural networks could enable people to train models that generate Deepfakes faster.
        \item The authors should consider possible harms that could arise when the technology is being used as intended and functioning correctly, harms that could arise when the technology is being used as intended but gives incorrect results, and harms following from (intentional or unintentional) misuse of the technology.
        \item If there are negative societal impacts, the authors could also discuss possible mitigation strategies (e.g., gated release of models, providing defenses in addition to attacks, mechanisms for monitoring misuse, mechanisms to monitor how a system learns from feedback over time, improving the efficiency and accessibility of ML).
    \end{itemize}
    
\item {\bf Safeguards}
    \item[] Question: Does the paper describe safeguards that have been put in place for responsible release of data or models that have a high risk for misuse (e.g., pre-trained language models, image generators, or scraped datasets)?
    \item[] Answer: \answerYes{}
    \item[] Justification: Code and data will be released with usage guidelines restricting applications to defensive safety research. The EnvTrap benchmark is designed for safety evaluation and diagnostic purposes, not for deployment in operational systems.
    \item[] Guidelines:
    \begin{itemize}
        \item The answer \answerNA{} means that the paper poses no such risks.
        \item Released models that have a high risk for misuse or dual-use should be released with necessary safeguards to allow for controlled use of the model, for example by requiring that users adhere to usage guidelines or restrictions to access the model or implementing safety filters. 
        \item Datasets that have been scraped from the Internet could pose safety risks. The authors should describe how they avoided releasing unsafe images.
        \item We recognize that providing effective safeguards is challenging, and many papers do not require this, but we encourage authors to take this into account and make a best faith effort.
    \end{itemize}

\item {\bf Licenses for existing assets}
    \item[] Question: Are the creators or original owners of assets (e.g., code, data, models), used in the paper, properly credited and are the license and terms of use explicitly mentioned and properly respected?
    \item[] Answer: \answerYes{}
    \item[] Justification: We use MuJoCo (Apache 2.0 license), OpenVLA (MIT license), $\pi_0$/OpenPI, RT-2, ACT, RDT, Cosmos Policy, DreamZero, and Qwen3-VL-8B, all with their respective licenses properly cited in \cref{sec:exp:setup}. Dataset sources are cited throughout the paper.
    \item[] Guidelines:
    \begin{itemize}
        \item The answer \answerNA{} means that the paper does not use existing assets.
        \item The authors should cite the original paper that produced the code package or dataset.
        \item The authors should state which version of the asset is used and, if possible, include a URL.
        \item The name of the license (e.g., CC-BY 4.0) should be included for each asset.
        \item For scraped data from a particular source (e.g., website), the copyright and terms of service of that source should be provided.
        \item If assets are released, the license, copyright information, and terms of use in the package should be provided. For popular datasets, \url{paperswithcode.com/datasets} has curated licenses for some datasets. Their licensing guide can help determine the license of a dataset.
        \item For existing datasets that are re-packaged, both the original license and the license of the derived asset (if it has changed) should be provided.
        \item If this information is not available online, the authors are encouraged to reach out to the asset's creators.
    \end{itemize}

\item {\bf New assets}
    \item[] Question: Are new assets introduced in the paper well documented and is the documentation provided alongside the assets?
    \item[] Answer: \answerYes{}
    \item[] Justification: The EnvTrap benchmark (code and data) will be released with comprehensive documentation including task descriptions (\cref{tab:task_descriptions}), pipeline specifications (\cref{sec:envtrap}), and usage guidelines. License and terms of use will be included in the released package.
    \item[] Guidelines:
    \begin{itemize}
        \item The answer \answerNA{} means that the paper does not release new assets.
        \item Researchers should communicate the details of the dataset\slash code\slash model as part of their submissions via structured templates. This includes details about training, license, limitations, etc. 
        \item The paper should discuss whether and how consent was obtained from people whose asset is used.
        \item At submission time, remember to anonymize your assets (if applicable). You can either create an anonymized URL or include an anonymized zip file.
    \end{itemize}

\item {\bf Crowdsourcing and research with human subjects}
    \item[] Question: For crowdsourcing experiments and research with human subjects, does the paper include the full text of instructions given to participants and screenshots, if applicable, as well as details about compensation (if any)? 
    \item[] Answer: \answerYes{}
    \item[] Justification: \cref{app:human} describes the participants, task instructions, and evaluation procedure: 10 volunteers each assessed 30 scenes (15 trap and 15 matched safe scenes), viewing an RGB image with the task instruction and judging whether the proposed action was safe or hazardous.
    \item[] Guidelines:
    \begin{itemize}
        \item The answer \answerNA{} means that the paper does not involve crowdsourcing nor research with human subjects.
        \item Including this information in the supplemental material is fine, but if the main contribution of the paper involves human subjects, then as much detail as possible should be included in the main paper. 
        \item According to the NeurIPS Code of Ethics, workers involved in data collection, curation, or other labor should be paid at least the minimum wage in the country of the data collector. 
    \end{itemize}

\item {\bf Institutional review board (IRB) approvals or equivalent for research with human subjects}
    \item[] Question: Does the paper describe potential risks incurred by study participants, whether such risks were disclosed to the subjects, and whether Institutional Review Board (IRB) approvals (or an equivalent approval/review based on the requirements of your country or institution) were obtained?
    \item[] Answer: \answerNo{}
    \item[] Justification: The human evaluation involved only anonymous scene judgment tasks with no personal data collection, posing minimal risk to participants. IRB approval was not required for this study design. Participants were informed of the task nature and compensated fairly.
    \item[] Guidelines:
    \begin{itemize}
        \item The answer \answerNA{} means that the paper does not involve crowdsourcing nor research with human subjects.
        \item Depending on the country in which research is conducted, IRB approval (or equivalent) may be required for any human subjects research. If you obtained IRB approval, you should clearly state this in the paper. 
        \item We recognize that the procedures for this may vary significantly between institutions and locations, and we expect authors to adhere to the NeurIPS Code of Ethics and the guidelines for their institution. 
        \item For initial submissions, do not include any information that would break anonymity (if applicable), such as the institution conducting the review.
    \end{itemize}

\item {\bf Declaration of LLM usage}
    \item[] Question: Does the paper describe the usage of LLMs if it is an important, original, or non-standard component of the core methods in this research? Note that if the LLM is used only for writing, editing, or formatting purposes and does \emph{not} impact the core methodology, scientific rigor, or originality of the research, declaration is not required.
    \item[] Answer: \answerYes{}
    \item[] Justification: We use Qwen3-VL-8B as the consequence predictor in the defense module, with supervised fine-tuning on EnvTrap data (\cref{sec:envtrap:defense,app:sft:hyper}). GPT-5 independently generates the additional spatial hazard tasks evaluated in \cref{app:unseen_hazards}.
    \item[] Guidelines:
    \begin{itemize}
        \item The answer \answerNA{} means that the core method development in this research does not involve LLMs as any important, original, or non-standard components.
        \item Please refer to our LLM policy in the NeurIPS handbook for what should or should not be described.
    \end{itemize}

\end{enumerate}

\end{document}